\documentclass[%
 reprint,
 amsmath,amssymb,
 aps,
]{revtex4-2}

\usepackage{graphicx}
\usepackage{dcolumn}
\usepackage{bm}
\usepackage{geometry}
\usepackage{xcolor}
\usepackage{mathtools}
\usepackage{amsmath}
\usepackage[normalem]{ulem}
\usepackage{amssymb}
\usepackage{babel}
\usepackage{tcolorbox}
\usepackage{comment}	
\usepackage{xr}
\usepackage{hyperref}
\newcommand{\hide}[1]{}
\usepackage{xcolor}
\definecolor{dark green}{rgb}{0.0, 0.5, 0.0}  

\begin{document}

\preprint{APS/123-QED}

\title{Hidden optical nonlinearities in linear spectra of quantum emitter arrays}

\author{Sricharan Raghavan-Chitra}
\author{Arghadip Koner
}
\author{Joel Yuen-Zhou}%
 \email{joelyuen@ucsd.edu}
\affiliation{%
 Department of Chemistry and Biochemistry, University of California San Diego, La Jolla, California 92093
}%
\date{\today}

\begin{abstract}
\noindent Classical optical frameworks such as the discrete dipole approximation (DDA) assume that the linear spectrum of coupled quantum emitters can be computed solely from the linear susceptibilities of individual constituents. However, recent polariton studies show that cavity linear response can encode nonlinear optical susceptibilities. Here, we demonstrate that this phenomenon is more general: emitter–emitter interactions allow nonlinearities of individual emitters to emerge in the linear response of arrays, without cavities or permutational symmetry. To illustrate this phenomenon, we show linear spectra for coupled heterodimers and linear chains, and demonstrate that Raman features of individual monomers show up as vibrational sidebands of collective resonances. Moreover, tuning Raman-type anharmonicities enables systematic control of spectral features, establishing a genuine quantum optical effect in molecular aggregates and quantum emitter arrays, which goes beyond mean-field descriptions in light-matter interactions.

\end{abstract}

\maketitle

\section{\label{sec:level1}Introduction}

\noindent The optical response of quantum emitter arrays~\cite{chang2018colloquium, asenjo2017exponential}, ranging from molecular aggregates~\cite{hestand2018expanded, patra2022vibronic, noblet2021diagrammatic, chen2009optical, wohlgemuth2020excitation, li2017probing, barford2017perspective, kocherzhenko2017absorption, deshmukh2019design} to solid-state \cite{aharonovich2016solid} and cavity-coupled platforms \cite{raimond2001manipulating, reiserer2015cavity}, lies at the heart of nanophotonics~\cite{myroshnychenko2008modelling, aizpurua2005optical} and quantum optics, with implications for light harvesting \cite{saikin2013photonics, fassioli2014photosynthetic}, sensing \cite{lee2021quantum, lee2016quantum}, and quantum technologies \cite{kimble2008quantum, saffman2010quantum}. A central objective in these settings is to understand how microscopic photophysical processes are encoded in macroscopic observables, particularly in linear optical spectra, which are routinely used to infer structure and dynamics. Traditionally, such spectra are interpreted as probes of single-excitation physics~\cite{MukamelBook}, providing access to transition energies, oscillator strengths, and coherence properties, while higher-order processes, such as Raman scattering and multiphoton interactions, are presumed to belong exclusively to the domain of nonlinear spectroscopy~\cite{MukamelBook}. This viewpoint is closely tied to a central tenet of classical linear optics~\cite{jackson1998classical}: the macroscopic response of a composite system is fully determined by the linear susceptibility $\chi^{(1)}(\omega)$ of its individual constituents. This assumption, in turn, underlies a broad class of mean-field frameworks, including the Discrete Dipole Approximation (DDA)~\cite{devoe1964optical, devoe1965optical, barr2021insights}, the Coherent Potential Approximation (CPA)~\cite{soven1967coherent, chenu2017construction}, and the Coherent Exciton Scattering (CES) approximation~\cite{eisfeld2002j}. This paradigm has proven remarkably fruitful, underpinning our understanding of electromagnetic scattering by arbitrarily shaped particles~\cite{yurkin2007discrete}, the properties of disordered alloys~\cite{soven1967coherent, velicky1969theory, soven1969contribution}, and the optical properties of quantum emitter arrays~\cite{briggs1970sum, briggs1971band, devoe1964optical, devoe1965optical}. \\ 

\noindent However, recent studies of light-matter systems~\cite{raghavanchitra2025high, koner2025hidden}, have called this classical picture into question by showing that the cavity-modified linear response of a molecular ensemble can encode contributions from nonlinear susceptibilities, specifically Raman-type contributions conventionally regarded as components of $\chi^{(3)}(\omega)$ and therefore, by the textbook account, invisible to linear spectroscopy. Two ingredients underlie this counterintuitive result: the permutational symmetry of identical molecules coupled to a common cavity mode~\cite{perez2023simulating, perez2025cut}, and the nature of the photophysical processes in the emitters that contribute to the overall linear response of the system. Subsequent work~\cite{raghavan2026permutationally} demonstrated that permutational symmetry alone, even in the absence of a cavity, is sufficient to generate analogous corrections in molecular aggregates, while simultaneously clarifying the precise regime in which CPA and CES are exact and where they break down. However, such systems are highly idealized and rarely realized in large molecular ensembles. Consequently, the scope of these results remains intrinsically tied to the presence of strict many-body permutational symmetry \cite{campos2021generalization, campos2022generalization}, thereby limiting their applicability to more general, structurally disordered emitter arrays. \\ 

\noindent In this article, we demonstrate that this phenomenon is far more general: emitter–emitter interactions enable intrinsic nonlinearities of individual emitters to manifest within the linear response of arrays, without requiring cavities or permutational symmetry. We illustrate this mechanism using the simplest non-symmetric coupled system: a heterodimer. Despite its minimal structure, this model captures essential physics relevant to a wide range of contemporary platforms, including coupled NV centers \cite{gaebel2006room}, lanthanide dimers \cite{hou2003synthesis}, quantum emitters in two-dimensional materials \cite{stern2019spectrally}, and molecular aggregates \cite{reppert2010lowest}. Although its linear response is routinely obtained via exact diagonalization or matrix inversion, we instead derive it through a Dyson expansion~\cite{MukamelBook}, treating inter-emitter coupling as the perturbation. This framework reveals the underlying photophysical processes governing the heterodimer’s linear response. By mapping the dynamics onto ladder diagrams~\cite{tokmakoff_nonlinear_notes}, we show that Raman-type processes contribute to the array's linear response, akin to phenomena in polaritonic and permutationally symmetric aggregates. For the Chl522–Chl520 heterodimer~\cite{reppert2010lowest}, Raman signatures of one emitter appear as sidebands to the absorption peak of the other, enabling extraction of Raman information from linear spectra. Extending this framework to linearly coupled quantum emitters, we identify regimes where such features were previously overlooked. Controlled tuning of Raman-type anharmonicities thus provides systematic control over spectral features, revealing a genuine quantum optical effect long sought in the field of molecular aggregates~\cite{duque2015classical, dutta2025quantum, briggs2011equivalence}.\\

\noindent The article is organized as follows. In Sec. II, we introduce the heterodimer Hamiltonian and derive its linear response by resumming terms in the Dyson expansion, with the detailed derivation presented in Supplementary Information section 3. Using these expressions, we simulate the linear spectrum of the Chl522–Chl520 heterodimer within a minimal three-level basis sufficient to capture the phenomenon of interest. In Sec. III, we extend the analysis to more physically relevant linear arrays of quantum emitters. Using parameter regimes previously explored in the molecular-aggregate literature, we identify overlooked Raman features in the linear response of J-aggregates. Our results show that aggregate linear absorption spectra inherently encode higher-order nonlinear susceptibilities, thereby enriching the structure–spectra relationship and revealing a chemically tunable handle for the rational design of quantum emitter arrays.

\section{Linear response of heterodimer}

\noindent In this section, we derive the linear absorption spectrum of an arbitrary heterodimer comprising monomers A and B. The linear absorption spectrum of a system driven by a weak incident laser field is given by~\cite{MukamelBook}:
\begin{equation}
\sigma(\omega)\propto -\Im \langle \hat{\mu}\,\hat{G}(\omega)\,\hat{\mu} \rangle,
\end{equation}
where $\hat{\mu}$ is the transition dipole moment operator coupling the system to the electric field of the incoming laser, and 
$ \hat{G}(\omega)=\frac{1}{\omega-\hat{H}+i0^{+}} $
is the retarded Green's function of the system, with $\hat{H}$ denoting the total Hamiltonian. The expectation value $\langle \cdot \rangle$ is taken with respect to the initial state of the system (typically, the ground state; see Eq.~\ref{Eq: linear_response_ground_state_starting_formula}). For a collection of heterodimers, the total dipole operator decomposes as $\hat{\mu}=\hat{\mu}_{A}+\hat{\mu}_{B}$, and the full Hamiltonian takes the form:
\begin{equation}
\hat{H}=\hat{H}_{A}+\hat{H}_{B}+\hat{H}_{I}.
\label{Eq: overall_Hamiltonian}
\end{equation}

\noindent Under the Born--Oppenheimer approximation, the molecular Hamiltonian of monomer A or B is:
\begin{equation}
\hat{H}_{A/B}=\hat{T}_{A/B}+V_{g}(q_{A/B})\,|g_{A/B}\rangle\langle g_{A/B}|+V_{e}(q_{A/B})\,|e_{A/B}\rangle\langle e_{A/B}|,
\end{equation}
where $\hat{T}$ is the nuclear kinetic energy operator, $V_{g/e}$ denote the ground and excited potential energy surfaces (PESs), and $q_{A/B}$ represents the set of intramolecular vibrational coordinates of monomer A or B. The interaction Hamiltonian captures the excitonic coupling between the two monomers~\cite{hestand2018expanded}:
\begin{equation}
\hat{H}_{I}=J\,|e_{A}\rangle\langle e_{B}|+\mathrm{h.c.},
\label{Eq: interaction_Hamiltonian}
\end{equation}
where $J$ is the excitonic coupling strength between the locally excited states of A and B.\\

\noindent To make the derivation tractable while retaining the essential physics, we employ a reduced basis in which each monomer is represented as a three-level (or $\Lambda$-type) system: two vibrational levels on the ground-state PES and one vibrational level on the excited-state PES. For simplicity, all Franck--Condon overlaps between vibronic levels are set to unity. While this truncation simplifies the algebra considerably, it captures the photophysical behavior of interest, and the results can be straightforwardly extended to an arbitrary vibronic basis. Within this reduced basis, the monomer Hamiltonians take the explicit form:
\begin{small}
\begin{align}
\hat{H}_{A}&=\omega_{g,0,A}\,|g_{A},0_{A}\rangle\langle g_{A},0_{A}| \nonumber\\
&\quad+\omega_{g,1,A}\,|g_{A},1_{A}\rangle\langle g_{A},1_{A}|+\omega_{e,0,A}\,|e_{A},0'_{A}\rangle\langle e_{A},0'_{A}|,\\
\hat{H}_{B}&=\omega_{g,0,B}\,|g_{B},0_{B}\rangle\langle g_{B},0_{B}| \nonumber\\
&\quad+\omega_{g,1,B}\,|g_{B},1_{B}\rangle\langle g_{B},1_{B}|+\omega_{e,0,B}\,|e_{B},0'_{B}\rangle\langle e_{B},0'_{B}|,\\
\hat{H}_{I}&=J\,|e_{A};g_{B}\rangle\langle g_{A};e_{B}|+\mathrm{h.c.}.
\end{align}
\end{small}

\noindent Here, $\omega_{g,0,A(B)}$ is the energy of the global ground state $|g_{A(B)},0_{A(B)}\rangle$ of monomer A(B), while $\omega_{g,1,A(B)}$ and $\omega_{e,0,A(B)}$ denote the energies of the vibrationally excited level on the ground-state PES and the vibrational ground level on the excited-state PES, respectively. The coupling $J$ mediates energy transfer between the local excitations $|e_{A};g_{B}\rangle$ and $|g_{A};e_{B}\rangle$, where the semicolon denotes a tensor product (i.e., $|e_{A};g_{B}\rangle\equiv|e_{A}\rangle\otimes|g_{B}\rangle$). At zero temperature, the linear absorption of this heterodimer system under a perturbative light source is:
\begin{equation}
\sigma(\omega)\propto -\Im\,\langle g_{A},0_{A};g_{B},0_{B}|\,\hat{\mu}\,\hat{G}(\omega)\,\hat{\mu}\,|g_{A},0_{A};g_{B},0_{B}\rangle,
\label{Eq: linear_response_ground_state_starting_formula}
\end{equation}
where the total dipole operator is $\hat{\mu}=\hat{\mu}_{A}+\hat{\mu}_{B}$, with:
$ \hat{\mu}_{A} = \mu_{M}^{A}\,|g_{A}\rangle\langle e_{A}|+\mathrm{h.c.}$, $ \hat{\mu}_{B} = \mu_{M}^{B}\,|g_{B}\rangle\langle e_{B}|+\mathrm{h.c.}, $
and $\mu_{M}^{A/B}$ is the transition dipole matrix element of monomer A/B. Expanding the expectation value yields four distinct contributions:
\begin{small}
\begin{align}
&\langle g_{A},0_{A};g_{B},0_{B}|\,\hat{\mu}\,\hat{G}(\omega)\,\hat{\mu}\,|g_{A},0_{A};g_{B},0_{B}\rangle \nonumber \\
&=|\mu_{M}^{A}|^{2}\langle e_{A},0'_{A};g_{B},0_{B}|\,\hat{G}(\omega)\,|e_{A},0'_{A};g_{B},0_{B}\rangle \nonumber \\
&\quad+|\mu_{M}^{A}||\mu_{M}^{B}|\langle e_{A},0'_{A};g_{B},0_{B}|\,\hat{G}(\omega)\,|g_{A},0_{A};e_{B},0'_{B}\rangle \nonumber \\
&\quad+|\mu_{M}^{A}||\mu_{M}^{B}|\langle g_{A},0_{A};e_{B},0'_{B}|\,\hat{G}(\omega)\,|e_{A},0'_{A};g_{B},0_{B}\rangle \nonumber \\
&\quad+|\mu_{M}^{B}|^{2}\langle g_{A},0_{A};e_{B},0'_{B}|\,\hat{G}(\omega)\,|g_{A},0_{A};e_{B},0'_{B}\rangle.
\label{Eq: linear_response_three_level_system_diagonal+off_diagonal_terms}
\end{align}
\end{small}

\noindent The first and fourth terms are diagonal matrix elements of the retarded Green's function, corresponding to the creation and destruction of the exciton in the same monomer. The second and third terms are the off-diagonal matrix elements of the retarded Green's function, corresponding to the creation of an exciton in one monomer and its destruction in the other. These elements and their diagramatic interpretations are computed explicitly in Section 1 of the Supplementary Information. We summarize here the final expressions obtained for various terms in Eq.~\ref{Eq: linear_response_three_level_system_diagonal+off_diagonal_terms}:

\begin{widetext}
\begin{align}
\langle e_{A},0'_{A};g_{B},0_{B}|\hat{G}(\omega)|e_{A},0'_{A};g_{B},0_{B}\rangle 
&= \frac{G_{e,A}\Big[1-J^{2}G_{e,A,v,B}(G_{e,B}+G_{v,A,e,B})\Big]}
{1-J^{2}(G_{e,A}+G_{e,A,v,B})(G_{e,B}+G_{v,A,e,B})}, \\
\langle g_{A},0_{A};e_{B},0'_{B}|\hat{G}(\omega)|g_{A},0_{A};e_{B},0'_{B}\rangle 
&= \frac{G_{e,B}\Big[1-J^{2}G_{v,A,e,B}(G_{e,A}+G_{e,A,v,B})\Big]}
{1-J^{2}(G_{e,B}+G_{v,A,e,B})(G_{e,A}+G_{e,A,v,B})}, \\
\langle e_{A},0'_{A};g_{B},0_{B}|\hat{G}(\omega)|g_{A},0_{A};e_{B},0'_{B}\rangle 
&= \langle g_{A},0_{A};e_{B},0'_{B}|\hat{G}(\omega)|e_{A},0'_{A};g_{B},0_{B}\rangle \nonumber \\
&= \frac{J G_{e,A} G_{e,B}}
{1-J^{2}(G_{e,A}+G_{e,A,v,B})(G_{e,B}+G_{v,A,e,B})}.
\end{align}
\end{widetext}

\noindent where
\begin{small}
\begin{align}
G_{e,A} &= \frac{1}{\omega-(\omega_{e,0,A}+\omega_{g,0,B})+i\frac{\gamma_{e,A}}{2}}, \\
G_{e,B} &= \frac{1}{\omega-(\omega_{e,0,B}+\omega_{g,0,A})+i\frac{\gamma_{e,B}}{2}}, \\
G_{v,A,e,B} &= \frac{1}{\omega-(\omega_{g,1,A}+\omega_{e,0,B})+i\frac{\gamma_{v,A}+\gamma_{e,B}}{2}}, \\
G_{e,A,v,B} &= \frac{1}{\omega-(\omega_{e,0,A}+\omega_{g,1,B})+i\frac{\gamma_{e,A}+\gamma_{v,B}}{2}}.
\end{align}
\end{small}

\noindent Here, $\gamma_{e,A}$ and $\gamma_{e,B}$ denote the phenomenological decay rates of the electronically excited states of monomers A and B, respectively, capturing their finite lifetimes. In addition, $\gamma_{v,A}$ and $\gamma_{v,B}$ denote the decay rates of the vibrationally excited levels on the ground-state potential energy surfaces of monomers A and B, respectively. Importantly, this framework can be straightforwardly extended to an arbitrary number of vibronic states on both the ground- and excited-state potential energy surfaces (PES). Consequently, one can compute the linear absorption spectrum of any heterodimer using this protocol, without any constraint on the value of $J$, i.e., the coupling need not be treated perturbatively. The only requirement is to parameterize the molecular dimer such that all parameters entering the Hamiltonian in Eq.~\ref{Eq: overall_Hamiltonian} are determined: specifically, the energies and wavefunctions of the vibronic states, and the dipole--dipole coupling strength $J$ appearing in Eq.~\ref{Eq: interaction_Hamiltonian}. Although the linear response expressions in Eq.~\ref{Eq: linear_response_three_level_system_diagonal+off_diagonal_terms} are exact, it is illuminating to explore certain perturbative regimes of this system, where $J$ serves as the order parameter, as we do below.\\

\begin{figure}[ht!]
    \includegraphics[width=1\linewidth]{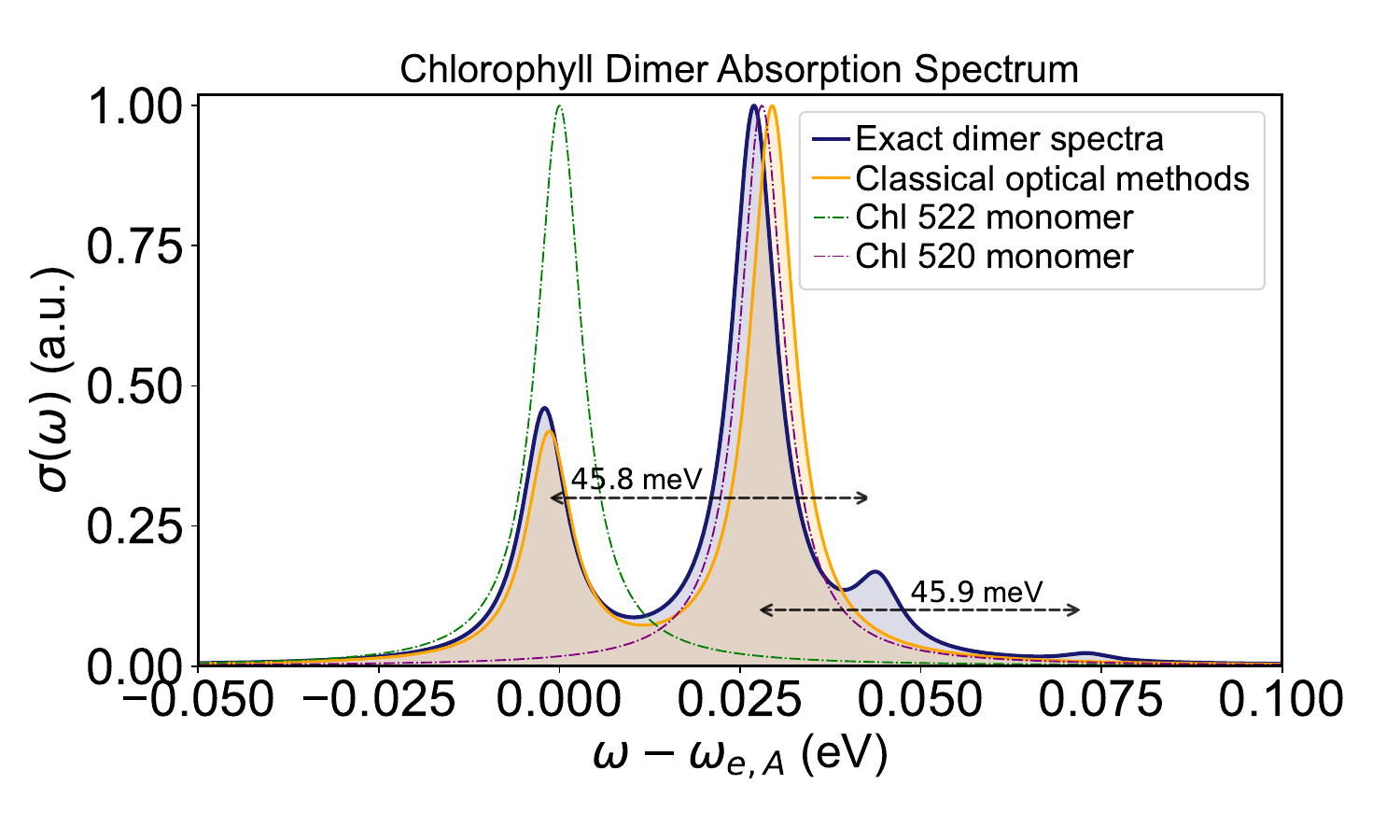}
    \caption{Absorption spectrum of the Chl522--Chl520 heterodimer computed using parameters from Ref.~\cite{reppert2010lowest} within a minimal three-level system basis that captures the essential physics of the phenomenon. Details of the simulations and explicit parameters are provided in Section~2 of the Supplementary Information. The blue curve shows the exact dimer spectrum obtained from Eq.~\ref{Eq: linear_response_three_level_system_diagonal+off_diagonal_terms}, while the orange curve corresponds to classical optical treatments (e.g., DDA/CPA/CES), which neglect contributions from ground-state vibrational excitations to the linear response by restricting each monomer to a single ground vibrational level. The red dash--dot curve shows the normalized monomer absorption of Chl522, and the black dash--dot curve shows that of Chl520. Notably, the exact spectrum (blue) exhibits additional sidebands that are absent in the classical mean-field descriptions. These features can be identified as Raman-type sidebands: for instance, the third peak at $\omega - \omega_{e,A} \approx 0.05\,\text{eV}$ is shifted from the Chl522 monomer peak by approximately the vibrational frequency of Chl520, while the fourth peak at $\omega - \omega_{e,A} \approx 0.075\,\text{eV}$ is similarly shifted from the Chl520 monomer peak by the vibrational frequency of Chl522. The theoretical derivations and perturbative analysis supporting these assignments are presented in Sections~1 and~2 of the Supplementary Information.}
\label{fig: dimer}
\end{figure}

\noindent We consider the linear response of a photosynthetically relevant heterodimer composed of Chl522 and Chl520, based on parameters summarized in Table 1 of the Supplementary Information, obtained from Ref.~\cite{reppert2010lowest}. The corresponding absorption spectrum is given in Fig.~\ref{fig: dimer}. Here, the absorption peak of Chl522 (A) monomer has an associated Raman (of Chl520 (B) monomer) sideband that is shifted by approximately the vibrational gap $\omega_{\nu,B}=\omega_{g,1,B}-\omega_{g,0,B}=0.043~\text{eV}$, and similarly, the absorption peak of Chl520 (B) monomer has an associated Raman (of Chl522 (A) monomer) sideband shifted by approximately $\omega_{\nu,A}=\omega_{g,1,A}-\omega_{g,0,A}=0.043~\text{eV}$. Thus, in this parameter regime, Raman-type information is directly encoded in the linear absorption spectrum of the heterodimer. This interpretation is consistent with the intuition developed from the ladder diagrams presented in Section~1 of the Supplementary Information: nonlinear contributions, specifically Raman-type processes associated with the third-order susceptibility, $\chi^{(3)}(\omega)$, are effectively encoded within the linear absorption spectrum of the dimer. This picture is further substantiated by the perturbative analysis presented in Section~2 of the Supplementary Information, where the intermolecular coupling $J$ in Eq.~\ref{Eq: linear_response_three_level_system_diagonal+off_diagonal_terms}serves as the perturbative expansion parameter.

\section{Linear Response in One-Dimensional Quantum Emitter Arrays}

\begin{figure}[ht!]
    \includegraphics[width=1\linewidth]{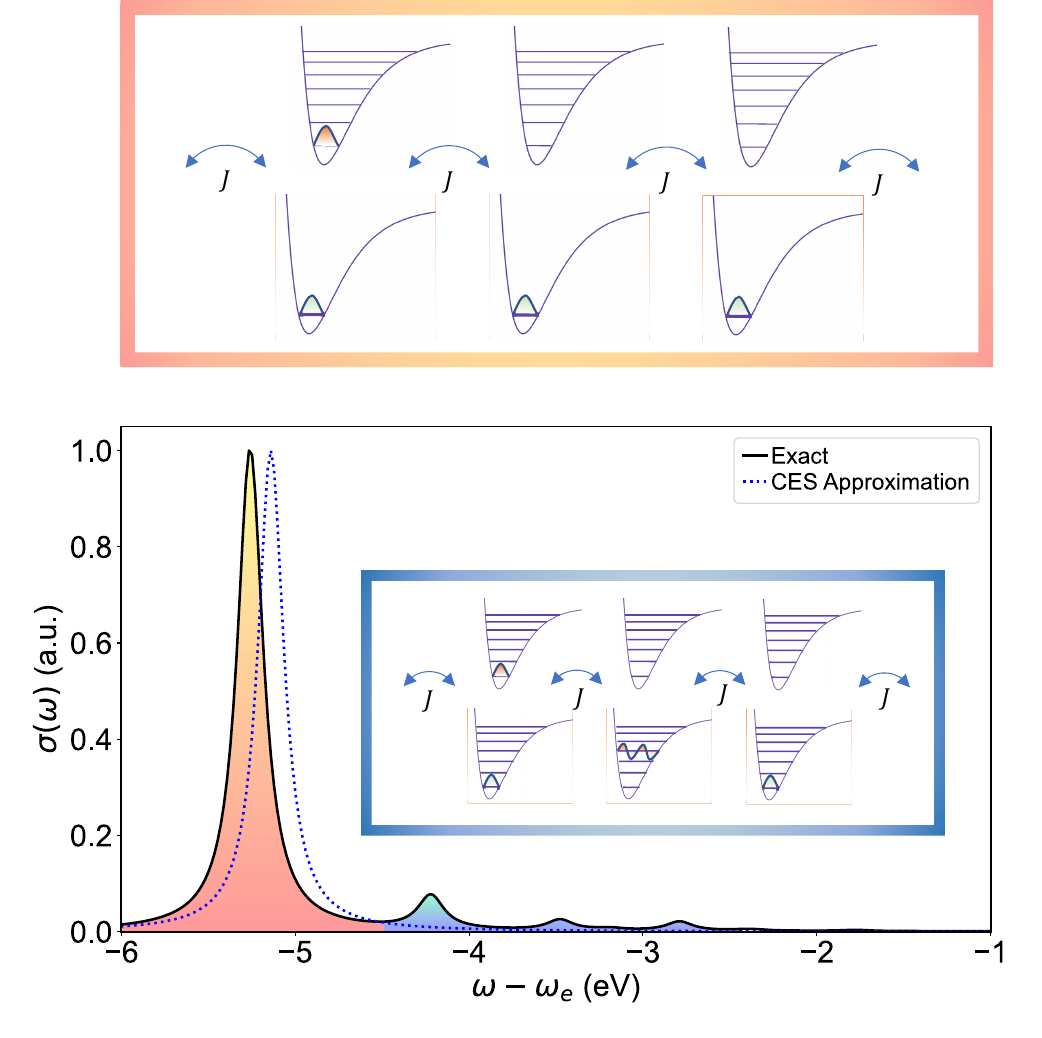}
    \caption{Linear absorption spectra of linearly coupled quantum emitter arrays comprising ten emitters computed by exact numerical diagonalization. The details of the model and the parameters are presented in Section 3 of the Supplementary Information. The DDA/CES/CPA captures the orange features in the exact spectra of the emitter arrays, analogous to those in Fig.~\ref{fig: dimer}, which physically arises from Rayleigh-type transitions, as schematically illustrated in the orange box. In contrast, spectral features such as the first blue-shifted peak, displaced from the orange peak by precisely one quantum of the ground-state vibrational frequency, are entirely absent within the DDA/CES/CPA approximations, in agreement with Fig.~\ref{fig: dimer} and originating from Raman-type transitions, as depicted in the blue schematic. Moreover, the spectra exhibit additional blue-shifted combination bands, approximately spaced by integer multiples of the vibrational frequency. We attribute this progression to higher-order photophysical pathways involving multiple monomers that host vibrational excitations in their ground electronic manifolds. Collectively, these features expose hidden nonlinear optical and vibronic correlations beyond the scope of CES and underscore the role of Raman-active processes in shaping the aggregate response. The simulation parameters are provided in Section 3 of the supplementary materials.}
\label{fig: linear_aggregate}
\end{figure}

\noindent In addition to the corrections observed in dimer spectra, these Raman processes also manifest as sidebands in the spectra of a strongly coupled quantum emitter array made up of ten emitters, as shown in Fig.~\ref{fig: linear_aggregate}. Here, we model the linear aggregate response using the standard shifted harmonic-oscillator framework~\cite{zheng2019non, zheng2019non, hestand2018expanded}. These features are consistent with those reported by Roden et al.~\cite{roden2008j}, who demonstrated, by varying the number of ground-state vibrational levels, which spectral features are captured by the CES approximation and which lie beyond its scope.\\

\noindent In this emitter array setup, our analysis of the dimer spectra provides a clear physical interpretation of these previously unexplained sidebands. We show that they originate from Raman-active vibrational modes of the monomer, which emerge in the linear spectra of the emitter arrays as sidebands. Additionally, the spectra display blue-shifted combination bands, spaced approximately by integer multiples of the vibrational frequency. We attribute this progression to higher-order photophysical pathways involving multiple monomers carrying vibrational excitations in their ground electronic states. Collectively, these insights reveal that linear absorption spectra of quantum emitter arrays encode richer vibrational information than previously appreciated, thereby deepening the structure–spectra relationship. Consequently, aggregate spectra can be exploited not only to probe excitonic couplings but also to infer ground-state vibrational fingerprints of the constituent monomers. Similarly, our results suggest a pathway for experimentally tuning aggregate spectral features through controlled modification of ground-state vibrational structure. This result is also consistent with the many-particle approximations~\cite{philpott1971theory} used by Spano et al.~\cite{spano2002absorption, spano2006excitons} to model aggregate absorption spectra. In particular, in certain strong-coupling regimes, these theories predict the emergence of side peaks (Fig. 17 (j) in ref:~\cite{hestand2018expanded}), which we interpret as Raman signatures (supplementary materials section 4).\\

\noindent In the present analysis, we have not explicitly incorporated inhomogeneous broadening. Nevertheless, our conclusions are expected to remain representative of experimentally relevant spectra~\cite{roden2008j, hestand2018expanded, spano2002absorption} in the regime $W>\omega_v>\sigma$, where $W$ is the exciton bandwidth, $\omega_v$ is the Raman vibrational frequency, and $\sigma$ is decay rate associated with the inhomogeneous broadening. This hierarchy defines an especially favorable window for observing Raman sidebands: the emitter array coupling is sufficiently strong compared to vibronic interactions to generate well-defined collective resonances, yet not so strong that the system reduces to the simple excitonic picture captured by Kasha’s theory (vibronic decoupling~\cite{hestand2018expanded}) ~\cite{kasha1963energy, kasha1965exciton}. Moreover, the dominant aggregate absorption peaks, being significantly shifted from their monomer counterparts, are known to exhibit reduced sensitivity to inhomogeneous broadening~\cite{houdre1996vacuum, roden2011anomalous}. Because the Raman features identified here appear in the vicinity of these robust aggregate resonances, we anticipate that disorder-induced broadening plays only a minor role in obscuring these spectral signatures within the regime of interest.

\section{conclusion}

\noindent In this work, by analyzing the linear response of coupled heterodimers and extending the framework to arrays of quantum emitters, we uncover the emergence of hidden optical nonlinearities encoded within the linear spectra of coupled systems. We show that these nonlinearities manifest as Raman-type signatures, closely analogous to those identified in cavity-coupled molecular ensembles and permutationally symmetric aggregates. In contrast to those settings, however, no permutational symmetry is required. Instead, the coexistence of Raman-active photophysical processes at the level of individual emitters, together with coherent intermolecular coupling, is sufficient for such nonlinear features to appear in the linear response of the composite system. Furthermore, the absence of permutational symmetry implies that these features are not subject to the $1/N$ suppression characteristic of symmetric systems \cite{schellenberger2026infinity, barberena2025generalized}, allowing them to persist at appreciable strength even in large ensembles. Finally, we demonstrate that such contributions are systematically missed by classical optical approaches, which, by construction, approximate the response solely in terms of the linear susceptibility of isolated monomers.\\

\noindent Importantly, we identify experimentally relevant parameter regimes, particularly within the chlorophyll heterodimer and molecular aggregate literature, that have been largely overlooked, yet exhibit clear Raman sideband signatures adjacent to the primary absorption peaks. Our analysis demonstrates that linear absorption spectra of molecular aggregates encode detailed ground-state vibrational information, thereby providing a principled route toward spectral engineering via controlled modification of monomer vibrational structure. Notably, these Raman-derived features are also captured by established many-body approaches, such as the two-particle approximation (TPA) \cite{spano2002absorption}, reinforcing their physical robustness. Finally, we anticipate that these signatures will remain experimentally observable in regimes of strong inter-emitter coupling, where disorder-resilient collective resonances preserve adjacent Raman sidebands.\\

\noindent At a broader level, Raman spectroscopy has been established as a cornerstone of molecular characterization, as it interrogates higher-order molecular susceptibilities that are fundamentally inaccessible to linear optical measurements. In this work, we demonstrate that such higher-order information of the monomers can naturally emerge as sidebands in the linear spectra of quantum emitter arrays. Remarkably, the underlying inter-emitter coupling, mediated by the electromagnetic vacuum, enables effective higher-order processes involving Raman-active vibrational modes, even in the presence of weak optical driving that cannot directly access these modes. This mechanism reveals a new pathway for exploiting inter-emitter interactions to probe individual emitter degrees of freedom that remain hidden to a weak laser field interacting with uncoupled emitters. Crucially, this physics lies beyond the scope of conventional classical-optics-based DDA/CES/CPA approaches, which model the linear spectra of coupled systems solely through the linear susceptibilities of their individual constituents.\\

\noindent Looking ahead, our work identifies specific regimes in quantum emitter array systems where informative Raman signatures emerge directly from the array linear optical response, thereby enabling experimental access to and characterization of the underlying monomeric units. Beyond spectroscopy, these results introduce a new and previously unexplored avenue for tunability in the design of efficient optoelectronic devices, namely through ground-state Raman-active vibrational modes that can directly influence the photophysical properties of the emitter array. More broadly, our results contribute to the growing body of genuinely quantum-mechanical phenomena in molecular aggregates by explicitly accounting for vibronic degrees of freedom. In the absence of such couplings, it has been argued that classical and quantum-coherent descriptions of emitter array dynamics become effectively indistinguishable~\cite{briggs2011equivalence}. In contrast, we show that the inclusion of molecular vibrations gives rise to distinctly quantum signatures that persist even in the linear response regime, traditionally regarded as adequately described by classical multichromophoric frameworks \cite{duque2015classical}. Because these effects originate from nonclassical inter-monomer correlations generated by inter-emitter coupling, we interpret the resulting Raman-induced spectral features as signatures of monomer-monomer entanglement, a connection that we plan to investigate quantitatively in future work. Altogether, our results open new directions in the study of quantum emitter arrays by demonstrating how inter-emitter interactions can harvest otherwise hidden optical nonlinearities, giving rise to emergent quantum effects and providing a new control knob for optimizing optoelectronic and quantum sensing performance.

\begin{acknowledgments}

This work was supported with a Camille Dreyfus Teacher-Scholar Award. S.R.-C. thanks Juan B. Pérez-Sánchez and Kai Schwennicke for useful discussions.

\end{acknowledgments}

\section*{AUTHOR DECLARATIONS}

\subsection*{Conflict of Interest}
The authors have no conflicts to disclose.

\subsection*{Data Availability}
Data sharing is not applicable to this article as no new data were created or analyzed in this study.
\bibliography{references}

\clearpage
\onecolumngrid

\setcounter{section}{0}
\setcounter{equation}{0}
\setcounter{figure}{0}
\setcounter{table}{0}

\setcounter{secnumdepth}{3}

\renewcommand{\thesection}{\arabic{section}}
\renewcommand{\thesubsection}{\thesection.\arabic{subsection}}
\renewcommand{\thesubsubsection}{\thesubsection.\arabic{subsubsection}}

\begin{center}
\textbf{\large Supplementary Information: Hidden nonlinearities in the linear spectra of coupled quantum emitters}\\[6pt]
Sricharan Raghavan-Chitra, Arghadip Koner, Joel Yuen-Zhou\\
\textit{Department of Chemistry and Biochemistry, University of California San Diego, La Jolla, California 92093}
\end{center}

\section{Diagonal and off-diagonal terms in in Eq. of the main text\label{sec:SI}}

In this section, we systematically evaluate both the diagonal and
off-diagonal matrix elements appearing on the right-hand side of Eq.~(9)
of the main text. The section is organized as follows: Sec.~\ref{subsec:diagonal_terms}
is devoted to the computation of the diagonal contributions, while
Sec.~\ref{subsec:Off-diagonal-terms} addresses the corresponding
off-diagonal contributions.

\subsection{Diagonal terms \label{subsec:diagonal_terms}}

In this subsection, we compute the diagonal terms in Eq. 9 of the
main text,
\begin{equation}
\langle e_{A},0'_{A};\,g_{B},0_{B}\,|\,\hat{G}(\omega)\,|\,e_{A},0'_{A};\,g_{B},0_{B}\rangle,
\end{equation}
via the Dyson expansion, by partitioning the full Hamiltonian as $\hat{H}=\hat{H}_{0}+\hat{V}$,
where $\hat{H}_{0}=\hat{H}_{A}+\hat{H}_{B}$ denotes the bare, non-interacting
Hamiltonian, and $\hat{V}=\hat{H}_{I}$ is the excitonic coupling,
treated as a perturbation. The Dyson expansion \cite{MukamelBook}
then yields: 
\begin{align}
\langle e_{A},0'_{A};\,g_{B},0_{B}\,|\,\hat{G}(\omega)\,|\,e_{A},0'_{A};\,g_{B},0_{B}\rangle & =\langle e_{A},0'_{A};\,g_{B},0_{B}\,|\,\hat{G}_{0}(\omega)\,|\,e_{A},0'_{A};\,g_{B},0_{B}\rangle\nonumber \\
 & \quad+\langle e_{A},0'_{A};\,g_{B},0_{B}\,|\,\hat{G}_{0}(\omega)\,\hat{V}\,\hat{G}_{0}(\omega)\,|\,e_{A},0'_{A};\,g_{B},0_{B}\rangle\nonumber \\
 & \quad+\langle e_{A},0'_{A};\,g_{B},0_{B}\,|\,\hat{G}_{0}(\omega)\,\hat{V}\,\hat{G}_{0}(\omega)\,\hat{V}\,\hat{G}_{0}(\omega)\,|\,e_{A},0'_{A};\,g_{B},0_{B}\rangle+\cdots,\label{eq:dyson_expansion}
\end{align}
where $\hat{G}_{0}(\omega)$ is the retarded Green's function of the
non-interacting system: 
\begin{equation}
\hat{G}_{0}(\omega)=\frac{1}{\omega-\hat{H}_{0}+i0^{+}}.
\end{equation}

The successive terms on the right-hand side of Eq.~\ref{eq:dyson_expansion}
are referred to as the zeroth-order, first-order, second-order contributions,
and so on, in the standard sense of the Dyson expansion in powers
of $\hat{V}$. We emphasize that this notion of ``order'' is distinct
from its usage in previous works from our research group based on
the CUT-E method (\cite{perez2023simulating,perez2025cut}). In the
following subsections, we systematically compute each of these terms
order by order.

\subsubsection{Zeroth-order term: $\langle e_{A},0'_{A};g_{B},0_{B}|\hat{G}_{0}(\omega)|e_{A},0'_{A};g_{B},0_{B}\rangle$}

The zeroth-order contribution is straightforwardly evaluated as: 
\begin{equation}
\langle e_{A},0'_{A};g_{B},0_{B}|\hat{G}_{0}(\omega)|e_{A},0'_{A};g_{B},0_{B}\rangle=\frac{1}{\omega-(\omega_{e,0,A}+\omega_{g,0,B})+i\frac{\gamma_{e,A}}{2}},\label{eq:zeroth_order}
\end{equation}
where $\gamma_{e,A}$ is the decay rate associated with the electronic
excited state of monomer A, introduced phenomenologically to account
for the finite lifetime of the excited state. For notational compactness,
we define: 
\begin{equation}
G_{e,A}\equiv\frac{1}{\omega-(\omega_{e,0,A}+\omega_{g,0,B})+i\frac{\gamma_{e,A}}{2}}.\label{eq:G_eA_def}
\end{equation}

\begin{figure}
\centering \includegraphics[width=0.4\linewidth]{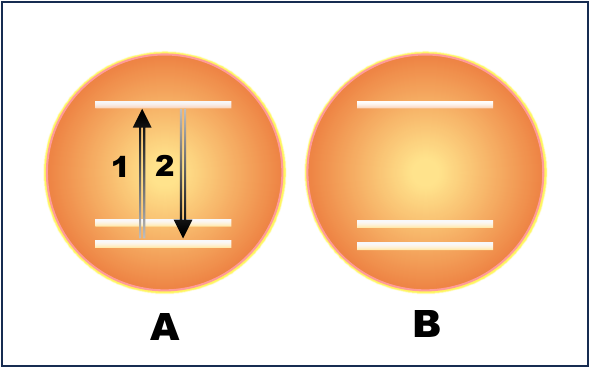}
\caption{The ladder diagram~\cite{tokmakoff_nonlinear_notes} of the heterodimer
illustrates the photophysical process in which the incident light
field interacts perturbatively with monomer~A alone. The double-line
arrows denote processes directly mediated by the incoming laser. The
numbers adjacent to each arrow indicate the sequence of interactions:
(1) excitation of monomer~A by the laser and (2) de-excitation of
monomer~A through a second interaction with the laser.}
\label{fig:ladder_diagram} 
\end{figure}

Physically, this zeroth-order term describes the free propagation
of the excitation on monomer A, in the absence of any inter-monomer
coupling. The corresponding photophysical process is depicted in the
ladder diagram in Fig.~\ref{fig:ladder_diagram}, where the double-line
arrow represents processes directly mediated by the incoming laser,
with no contribution from the perturbation $\hat{V}$.

\subsubsection{First-order term: $\langle e_{A},0'_{A};g_{B},0_{B}|\hat{G}_{0}\hat{V}\hat{G}_{0}|e_{A},0'_{A};g_{B},0_{B}\rangle$}

The first-order contribution to the Dyson expansion evaluates as:
\begin{equation}
\langle e_{A},0'_{A};g_{B},0_{B}|\hat{G}_{0}\hat{V}\hat{G}_{0}|e_{A},0'_{A};g_{B},0_{B}\rangle=G_{e,A}^{2}\,\langle e_{A},0'_{A};g_{B},0_{B}|\hat{V}|e_{A},0'_{A};g_{B},0_{B}\rangle=0.\label{eq:first_order_zero}
\end{equation}
This vanishes because the interaction Hamiltonian $\hat{V}=H_{I}$
couples states of the form $|e_{A};g_{B}\rangle\leftrightarrow|g_{A};e_{B}\rangle$,
and therefore has no diagonal matrix elements in this basis. By the
same argument, all odd-order terms in the Dyson expansion of Eq.~(\ref{eq:dyson_expansion})
vanish identically, since each such term necessarily contains an odd
power of $\hat{V}$ sandwiched between the same bra and ket, which
cannot be connected by an odd number of excitation-transfer operations.

\subsubsection{Second-order term: $\langle e_{A},0'_{A};g_{B},0_{B}|\hat{G}_{0}\hat{V}\hat{G}_{0}\hat{V}\hat{G}_{0}|e_{A},0'_{A};g_{B},0_{B}\rangle$\label{subsec:second_order_diag}}

\begin{figure}
\centering\includegraphics[scale=0.75]{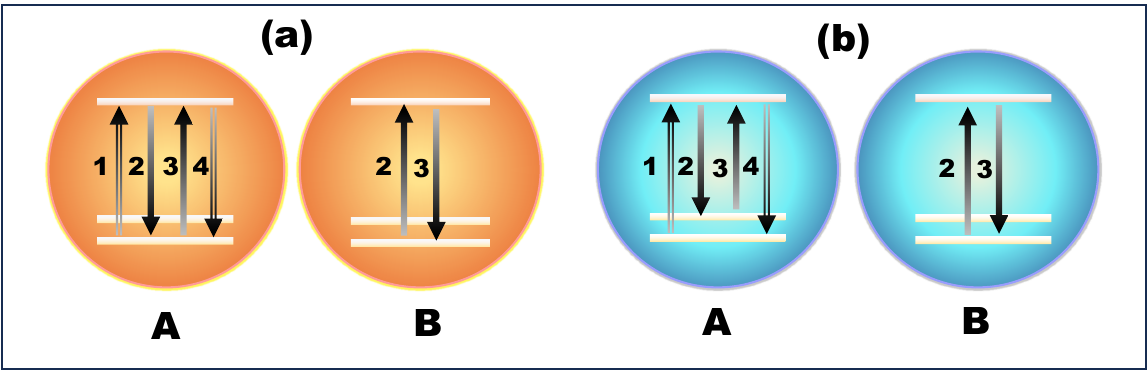}\caption{The ladder diagram~\cite{tokmakoff_nonlinear_notes} corresponding
to the second-order term in Eq.~{[}eq:dysonexpansion{]} illustrates
the case where the light field interacts perturbatively with monomer~A
alone. Both diagrams involve two successive dipole-dipole interactions,
through which the excitation is transferred to monomer~B and subsequently
returned to monomer~A. Panel~(a) captures a purely Rayleigh-type
photophysical process, with no vibrational excitation in the ground
electronic state, whereas panel~(b) includes a Raman-type contribution
in which vibrational coherence is generated in monomer~A. The double-line
arrows denote processes directly mediated by the incoming laser, while
the solid arrows represent dipole-dipole ($J$-coupling) mediated
processes. The numbers adjacent to each arrow indicate the sequence
of interactions: (1) excitation of monomer~A by the laser, (2--3)
exciton transfer from A to B via dipole-dipole coupling, and (4) de-excitation
of monomer~A through a second interaction with the laser.}
\label{fig:ladder_second_order} 
\end{figure}

The leading non-vanishing contribution in the Dyson expansion that
involves the coupling $\hat{V}$ arises at second order. Evaluating
explicitly, we find: 
\begin{equation}
\langle e_{A},0'_{A};\,g_{B},0_{B}\,|\,\hat{G}_{0}(\omega)\,\hat{V}\,\hat{G}_{0}(\omega)\,\hat{V}\,\hat{G}_{0}(\omega)\,|\,e_{A},0'_{A};\,g_{B},0_{B}\rangle=J^{2}\,G_{e,A}^{2}\left(G_{e,B}+G_{v,A,e,B}\right),\label{eq:second_order-1}
\end{equation}
where we have defined: 
\begin{align}
G_{e,B} & \equiv\frac{1}{\omega-\left(\omega_{e,0,B}+\omega_{g,0,A}\right)+i\,\dfrac{\gamma_{e,B}}{2}},\label{eq:GeB}\\[10pt]
G_{v,A,e,B} & \equiv\frac{1}{\omega-\left(\omega_{e,0,B}+\omega_{g,1,A}\right)+i\,\dfrac{\gamma_{e,B}+\gamma_{v,A}}{2}}.\label{eq:GvAeB}
\end{align}
Here, $\gamma_{e,B}$ is the decay rate of the electronic excited
state of monomer $B$, and $\gamma_{v,A}$ is the decay rate of the
vibrationally excited level on the ground-state potential energy surface
(PES) of monomer $A$.

The physical content of this result is particularly illuminating.
The factor $J^{2}$ reflects the second-order nature of the perturbation
$\hat{V}$. The factor $G_{e,A}^{2}$ accounts for two free propagations
of the system in the state $|e_{A},0'_{A};\,g_{B},0_{B}\rangle$ ---
once immediately following the first laser interaction, and once before
the second. Notably, since this term is second order in $\hat{V}$,
monomer $A$ exchanges the exciton twice with monomer $B$, returning
to $A$ before the second laser interaction takes place. During this
exciton exchange, two distinct pathways are available. First, the
exciton is exchanged with $B$ without depositing any phonon in the
ground state of $A$ --- this is referred to as the \textbf{Rayleigh
process} (depicted in Fig.~\ref{fig:ladder_second_order}(a)). Second,
the exciton is exchanged with $B$ while leaving a phonon in the ground
state of $A$ --- this is referred to as the \textbf{Raman process}
(depicted in Fig.~\ref{fig:ladder_second_order}(b)). These two pathways
are captured in the factor $(G_{e,B}+G_{v,A,e,B})$.

Remarkably, we already observe that a process known to capture the
nonlinear susceptibility of the monomer is contributing to the \textit{linear}
response of the dimer. This emerges as a recurring theme throughout
the derivation: at each successive even order of perturbation theory,
additional Raman-type processes appear, progressively encoding the
nonlinear response of the individual monomers into the linear response
of the dimer.

\subsubsection{Fourth-order term: $\langle e_{A},0'_{A};\,g_{B},0_{B}\,|\,\hat{G}_{0}(\omega)\,\hat{V}\,\hat{G}_{0}(\omega)\,\hat{V}\,\hat{G}_{0}(\omega)\,\hat{V}\,\hat{G}_{0}(\omega)\,\hat{V}\,\hat{G}_{0}(\omega)\,|\,e_{A},0'_{A};\,g_{B},0_{B}\rangle$\label{subsec:fourth_order_diagonal_term}}

\begin{figure}[ht!]
\centering{}\includegraphics[width=1\linewidth]{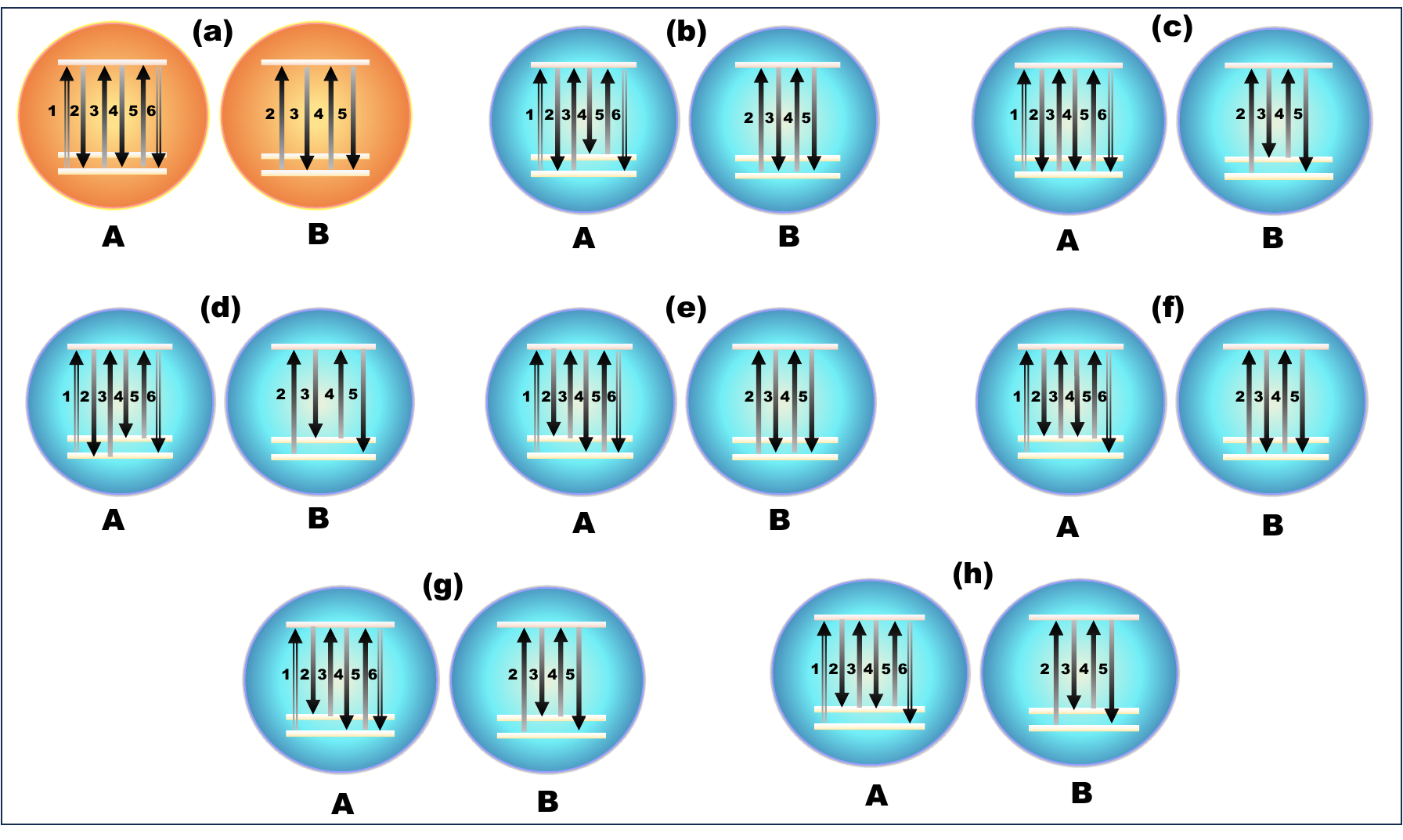}
\caption{All possible ladder diagrams~\cite{tokmakoff_nonlinear_notes} are
shown corresponding to fourth order in the interaction between the
monomers. Further, these diagrams correspond to the component where
the light source interacts only with monomer A. Only Rayleigh-type
photophysical processes takes place in the orange diagram (a), whereas,
Raman-type processes also takes place in the blue diagrams (b-h).
Here, the double line arrows correspond to direct processes due to
the laser and the solid line arrows correspond to processes due to
the dipole-dipole interaction ($J$- coupling). The numbers adjacent
to each arrow indicate the sequence of interactions: (1) excitation
of monomer~A by the laser, (2--5) exciton transfer from A to B via
dipole-dipole coupling, and (6) de-excitation of monomer~A through
a second interaction with the laser.}
\label{fig: ladder_diag_fourth-1}
\end{figure}

The diagrams shown in Fig.~\ref{fig:ladder_second_order} serve two
purposes. First, they make explicit the underlying photophysical processes
induced by the laser interactions, represented by double-line arrows,
and by the dipole--dipole interactions, represented by solid arrows.
This representation is particularly useful for identifying pathways
that encode Raman-type information. Second, the diagrams provide a
systematic bookkeeping device for generating all possible processes
at a given order in perturbation theory. There is a one-to-one correspondence
between these ladder diagrams and Double-Sided Feynman Diagrams, as
explored in ref: \cite{koner2025hidden}, but we believe it might
be simpler for the reader to understand the former in case they are
not familiar with the latter.

In particular, during exciton exchange between the two monomers, two
possibilities always arise: either the monomer that loses the exciton
is left without a vibrational excitation, or it is left with a vibrational
excitation. Consequently, the diagrammatic construction offers a transparent
way of enumerating all allowed contributions at a fixed perturbative
order.

This procedure is illustrated for the fourth-order contribution in
Eq.~\ref{eq:dyson_expansion}, 
\begin{equation}
\langle e_{A},0'_{A};\,g_{B},0_{B}\,|\,\hat{G}_{0}(\omega)\,\hat{V}\,\hat{G}_{0}(\omega)\,\hat{V}\,\hat{G}_{0}(\omega)\,\hat{V}\,\hat{G}_{0}(\omega)\,\hat{V}\,\hat{G}_{0}(\omega)\,|\,e_{A},0'_{A};\,g_{B},0_{B}\rangle\label{eq:fourth_order_diag_term}
\end{equation}
in Fig.~\ref{fig: ladder_diag_fourth-1}. In the present minimal
model, the role of the interaction operator $\hat{V}$ is solely to
exchange the exciton between the two monomers, while contributing
a factor of $J$ at each application. All energy denominators, and
hence all poles of the response, originate from the free propagator
$\hat{G}_{0}(\omega)$. Thus, once a diagram is specified, its analytic
expression can be straightforwardly obtained by reading off the sequence
of intermediate states generated after each interaction and writing
down the corresponding free propagators in the same order.

The numbers attached to the arrows in Fig.~\ref{fig: ladder_diag_fourth-1}
indicate the sequence of interactions, to be read from left to right
in Eq.~\ref{eq:fourth_order_diag_term}. It is important to note
that the bra and ket appearing in Eq.~\ref{eq:fourth_order_diag_term}
already correspond to the state produced after the initial laser interaction,
which is depicted by the double-line arrows in Fig.~\ref{fig: ladder_diag_fourth-1}.
Accordingly, Fig.~\ref{fig: ladder_diag_fourth-1} contains all possible
diagrams generated by the repeated action of $\hat{V}$, with the
initial and final laser interactions held fixed. In the present case,
both of these laser interactions occur on monomer $A$.

To illustrate the protocol, consider a specific diagram, Fig.~\ref{fig: ladder_diag_fourth-1}(b).
Its analytic contribution is obtained by recording the free-propagation
factor associated with each intermediate state after every interaction,
including the first laser-induced excitation. For the diagram under
consideration, this yields 
\begin{equation}
G_{e,A}\,J\,G_{e,B}\,J\,G_{e,A}\,J\,G_{e,B,v,A}\,J\,G_{e,A},
\end{equation}
where the ordering of the propagators follows directly the ordering
of the arrows in the diagram.

Repeating this procedure for all fourth-order diagrams and summing
the resulting contributions gives 
\begin{equation}
\begin{aligned} & \langle e_{A},0'_{A};\,g_{B},0_{B}\,|\,\hat{G}_{0}(\omega)\,\hat{V}\,\hat{G}_{0}(\omega)\,\hat{V}\,\hat{G}_{0}(\omega)\,\hat{V}\,\hat{G}_{0}(\omega)\,\hat{V}\,\hat{G}_{0}(\omega)\,|\,e_{A},0'_{A};\,g_{B},0_{B}\rangle\\
 & =J^{2}G_{e,A}^{2}(G_{e,B}+G_{v,A,e,B})\,J^{2}(G_{e,A}+G_{e,A,v,B})(G_{e,B}+G_{v,A,e,B}).
\end{aligned}
\label{eq:fourth_order_result}
\end{equation}

\noindent One can rationalize this structure as follows.The prefactor
$J^{2}G_{e,A}^{2}\left(G_{e,B}+G_{v,A,e,B}\right)$ has already been
established in the analysis of the second-order contribution (Sec.~\ref{subsec:second_order_diag}),
where it arises due to the constraint that the laser interacts exclusively
with monomer $A$. In contrast, the two additional applications of
the interaction operator $\hat{V}$, corresponding to arrows 3 and
4 in Fig.~\ref{fig: ladder_diag_fourth-1}(b), are unconstrained
and free to generate every intermediate state compatible with exciton
exchange between the two monomers within the present minimal model.
As a result, the remaining contributions organize into the factor
$J^{2}\left(G_{e,A}+G_{e,A,v,B}\right)\left(G_{e,B}+G_{v,A,e,B}\right)$,
where the prefactor $J^{2}$ arises from the two additional applications
of $\hat{V}$. The term $\left(G_{e,A}+G_{e,A,v,B}\right)$ enumerates
all possible configurations following an exciton transfer from monomer
$B$ to monomer $A$, while $\left(G_{e,B}+G_{v,A,e,B}\right)$ collects
the corresponding configurations following an exciton transfer from
monomer $A$ to monomer $B$.

\subsubsection{Resummation}

One can now recognize the emerging structure of the perturbation series.
In particular, by following the same diagrammatic logic used at fourth
order, one finds that the sixth-order contribution takes the form
\[
J^{2}G_{e,A}^{2}(G_{e,B}+G_{v,A,e,B})\big[J^{2}(G_{e,A}+G_{e,A,v,B})(G_{e,B}+G_{v,A,e,B})\big]^{2}.
\]
The key point is that each additional pair of interactions with $\hat{V}$
contributes the same building block, 
\[
J^{2}(G_{e,A}+G_{e,A,v,B})(G_{e,B}+G_{v,A,e,B}),
\]
which accounts for all allowed intermediate excitonic and vibronic
configurations generated by one further round-trip of the exciton
between monomers $A$ and $B$. Thus, the perturbation series assumes
a geometric structure, allowing all orders in $\hat{V}$ to be resummed
exactly.

As a result, the exact response corresponding to $\langle e_{A},0'_{A};\,g_{B},0_{B}\,|\,\hat{G}(\omega)\,|\,e_{A},0'_{A};\,g_{B},0_{B}\rangle$
(in Eq. 9 of the main text ) can be written as 
\[
\langle e_{A},0'_{A};\,g_{B},0_{B}\,|\,\hat{G}(\omega)\,|\,e_{A},0'_{A};\,g_{B},0_{B}\rangle=G_{e,A}+J^{2}G_{e,A}^{2}(G_{e,B}+G_{v,A,e,B})\sum_{n=0}^{\infty}\Big[J^{2}(G_{e,A}+G_{e,A,v,B})(G_{e,B}+G_{v,A,e,B})\Big]^{n}.
\]
Since this is a geometric series, it may be resummed immediately to
give 
\[
\langle e_{A},0'_{A};\,g_{B},0_{B}\,|\,\hat{G}(\omega)\,|\,e_{A},0'_{A};\,g_{B},0_{B}\rangle=G_{e,A}+\frac{J^{2}G_{e,A}^{2}(G_{e,B}+G_{v,A,e,B})}{1-J^{2}(G_{e,A}+G_{e,A,v,B})(G_{e,B}+G_{v,A,e,B})}.
\]
Upon combining terms, this expression can be recast in the compact
form 
\[
\langle e_{A},0'_{A};\,g_{B},0_{B}\,|\,\hat{G}(\omega)\,|\,e_{A},0'_{A};\,g_{B},0_{B}\rangle=\frac{G_{e,A}\Big[1-J^{2}G_{e,A,v,B}(G_{e,B}+G_{v,A,e,B})\Big]}{1-J^{2}(G_{e,A}+G_{e,A,v,B})(G_{e,B}+G_{v,A,e,B})}.
\]

By symmetry under interchange of the monomer labels $A\leftrightarrow B$,
one immediately obtains the corresponding diagonal matrix element
for the configuration in which the exciton resides on monomer $B$(refer
Eq. 9 of the main text), namely 
\[
\langle e_{B},0'_{B};\,g_{A},0_{A}\,|\,\hat{G}(\omega)\,|\,e_{B},0'_{B};\,g_{A},0_{A}\rangle=\frac{G_{e,B}\Big[1-J^{2}G_{v,A,e,B}(G_{e,A}+G_{e,A,v,B})\Big]}{1-J^{2}(G_{e,B}+G_{v,A,e,B})(G_{e,A}+G_{e,A,v,B})}.
\]

\subsection{Off diagonal terms in Eq. 9 of the main text\label{subsec:Off-diagonal-terms}}

Following the procedure for the diagonal elements mentioned in Sec.
\ref{subsec:diagonal_terms}, we can compute the exact expression
for the off diagonal elements, namely, $\langle e_{A},0'_{A};g_{B},0_{B}|\,\hat{G}(\omega)\,|g_{A},0_{A};e_{B},0'_{B}\rangle$
using dyson expansion and ladder diagrams. Thus, 
\begin{align}
\langle e_{A},0'_{A};g_{B},0_{B}|G(\omega)\left|g_{A},0_{A};e_{B},0'_{B}\right\rangle  & =\langle e_{A},0'_{A};g_{B},0_{B}|G_{0}\left|g_{A},0_{A};e_{B},0'_{B}\right\rangle \nonumber \\
 & \quad+\langle e_{A},0'_{A};g_{B},0_{B}|G_{0}\hat{V}G_{0}\left|g_{A},0_{A};e_{B},0'_{B}\right\rangle \nonumber \\
 & \quad+\langle e_{A},0'_{A};g_{B},0_{B}|G_{0}\hat{V}G_{0}\hat{V}G_{0}\left|g_{A},0_{A};e_{B},0'_{B}\right\rangle \nonumber \\
 & \quad+\dots
\end{align}

\subsubsection{\textcolor{black}{Zeroth order:} $\langle e_{A},0'_{A};g_{B},0_{B}|G_{0}\left|g_{A},0_{A};e_{B},0'_{B}\right\rangle $}

Here, the zeroth-order term is 
\[
\langle e_{A},0'_{A};g_{B},0_{B}|G_{0}|g_{A},0_{A};e_{B},0'_{B}\rangle=0,
\]
and, more generally, all even-order terms vanish. The reason is that
each application of the perturbation exchanges the exciton between
the two monomers. Therefore, after an even number of such interactions,
the exciton necessarily returns to the monomer on which it was initially
localized. As a result, even-order processes cannot contribute to
off-diagonal matrix elements in this basis.

\begin{figure}[ht!]
\centering{}\includegraphics[width=0.4\linewidth]{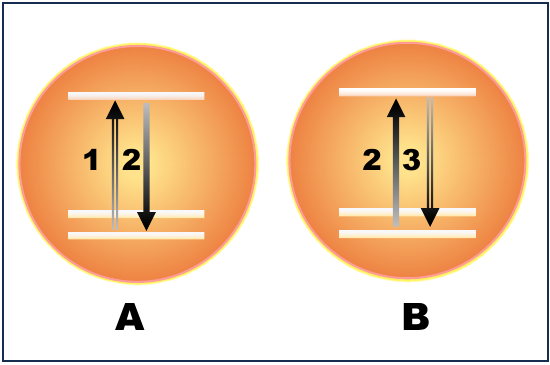}
\caption{The ladder diagram \cite{tokmakoff_nonlinear_notes} represents the
pathway in which the light field excites monomer~A and subsequently
de-excites monomer~B. This diagram corresponds to a first-order interaction
process. The double-line arrows denote direct interactions with the
laser field, while the solid arrows represent dipole-dipole ($J$-coupling)
mediated processes. The numbers adjacent to each arrow indicate the
sequence of interactions: (1) excitation of monomer~A by the laser,
(2) exciton transfer from A to B via dipole-dipole coupling, and (3)
de-excitation of monomer~B through a second interaction with the
laser.}
\label{fig: ladder_off_diag_first-1}
\end{figure}

\subsubsection{\textcolor{black}{First order:} $\langle e_{A},0'_{A};g_{B},0_{B}|G_{0}\hat{V}G_{0}\left|g_{A},0_{A};e_{B},0'_{B}\right\rangle $}

The first-order term is 
\[
\langle e_{A},0'_{A};g_{B},0_{B}|G_{0}\hat{V}G_{0}|g_{A},0_{A};e_{B},0'_{B}\rangle=JG_{e,A}G_{e,B}.
\]
The corresponding photophysical process is illustrated by the ladder
diagrams in Fig.~\ref{fig: ladder_off_diag_first-1}. Here, the diagrammatic
intuition developed in Sec.~\ref{subsec:fourth_order_diagonal_term}
is borne out transparently: the factor $J$ arises from the single
application of the perturbation $\hat{V}$, while $G_{e,A}$ and $G_{e,B}$
describe the free propagation of the system before and after the exciton
is exchanged from monomer $B$ to monomer $A$, respectively.

\begin{figure}[ht!]
\centering{}\includegraphics[width=0.75\linewidth]{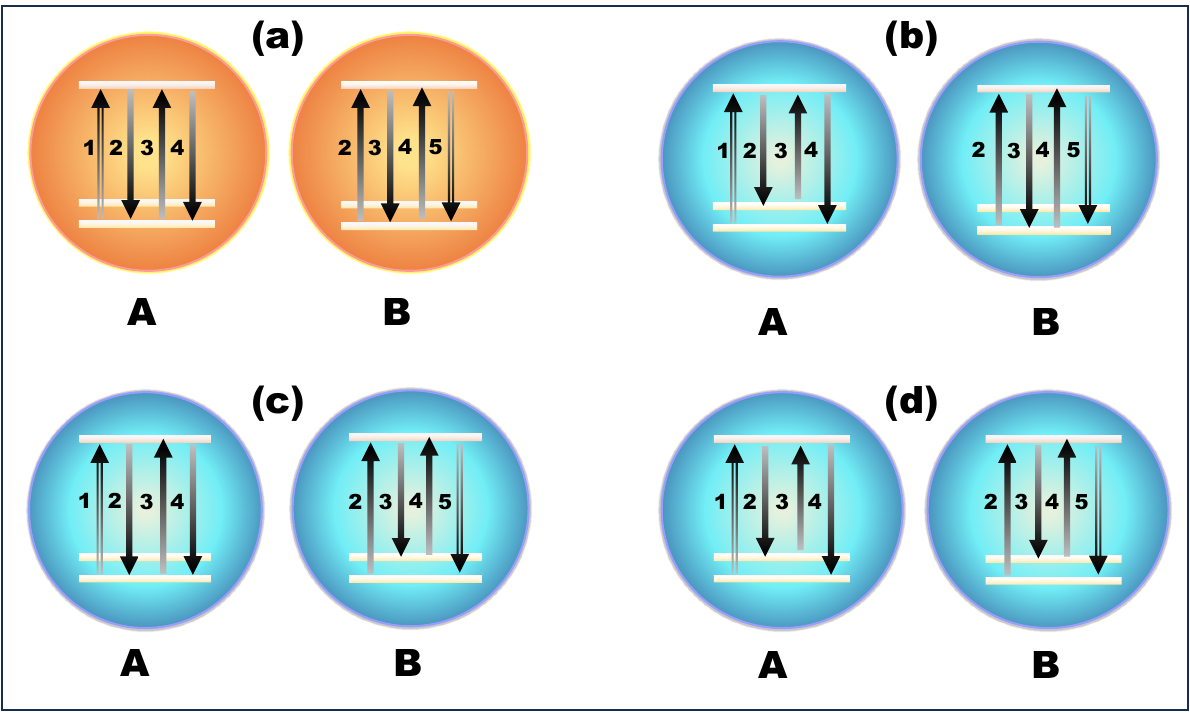}
\caption{The ladder diagram \cite{tokmakoff_nonlinear_notes} represents the
pathway in which the light field excites monomer~A and subsequently
de-excites monomer~B. This diagram corresponds to a third-order interaction
process. The orange diagram~(a) captures only Rayleigh-type processes,
whereas the blue diagrams~(b--d) additionally include Raman-type
processes. The double-line arrows denote direct interactions with
the laser field, while the solid arrows represent dipole-dipole ($J$-coupling)
mediated processes. The numbers adjacent to each arrow indicate the
sequence of interactions: (1) excitation of monomer~A by the laser,
(2--4) exciton transfer from A to B via dipole-dipole coupling, and
(5) de-excitation of monomer~B through a second interaction with
the laser.}
\label{fig: ladder_off_diag_third-1}
\end{figure}

\subsubsection{Resummation}

The photophysical processes contributing to the third-order term are
illustrated by the ladder diagrams in Fig.~\ref{fig: ladder_off_diag_third-1}.
Summing these diagrams yields 
\begin{equation}
JG_{e,A}G_{e,B}\,J^{2}(G_{e,A}+G_{e,A,v,B})(G_{e,B}+G_{v,A,e,B}).\label{eq: third_order_off_diagonal_term}
\end{equation}
The appearance of the additional factor 
\[
J^{2}(G_{e,A}+G_{e,A,v,B})(G_{e,B}+G_{v,A,e,B})
\]
is not incidental. It is precisely the same building block that appeared
in the higher-order contributions to the diagonal matrix elements
discussed in Sec.~\ref{subsec:fourth_order_diagonal_term}. The reason
is identical: once the initial and final excitonic configurations
are fixed, the two additional insertions of the interaction operator
$\hat{V}$ are free to generate all intermediate states allowed by
exciton exchange within the present minimal model. 

Now, summing the all non-vanishing odd order peturbative terms lead
to 
\begin{align}
\langle e_{A},0'_{A};g_{B},0_{B}|G(\omega)\left|g_{A},0_{A};e_{B},0'_{B}\right\rangle  & =JG_{e,A}G_{e,B}\sum_{n=0}^{\infty}\Big[J^{2}(G_{e,A}+G_{e,A,v,B})(G_{e,B}+G_{v,A,e,B})\Big]^{n}\nonumber \\
 & =\frac{JG_{e,A}G_{e,B}}{1-J^{2}(G_{e,A}+G_{e,A,v,B})(G_{e,B}+G_{v,A,e,B})}.
\end{align}

\noindent Again, by symmetry, 
\begin{equation}
\langle g_{A},0{}_{A};e_{B},0'_{B}|G(\omega)\left|e_{A},0'_{A};g_{B},0_{B}\right\rangle =\frac{JG_{e,A}G_{e,B}}{1-J^{2}(G_{e,A}+G_{e,A,v,B})(G_{e,B}+G_{v,A,e,B})}.
\end{equation}

\subsection{Final expression for the absorption spectrum of the dimer}

\noindent Thus, the final expression for the absorption spectrum of
the molecular dimer is, 
\begin{align}
\langle g_{A},0_{A};g_{B},0_{B}|\hat{\mu}\hat{G}(\omega)\hat{\mu}\left|g_{A},0_{A};g_{B},0_{B}\right\rangle  & =|\mu_{M}^{A}|^{2}\langle e_{A},0'_{A};g_{B},0_{B}|\,\hat{G}(\omega)\,|e_{A},0'_{A};g_{B},0_{B}\rangle\nonumber \\
 & \quad+|\mu_{M}^{A}||\mu_{M}^{B}|\langle e_{A},0'_{A};g_{B},0_{B}|\,\hat{G}(\omega)\,|g_{A},0_{A};e_{B},0'_{B}\rangle\nonumber \\
 & \quad+|\mu_{M}^{A}||\mu_{M}^{B}|\langle g_{A},0_{A};e_{B},0'_{B}|\,\hat{G}(\omega)\,|e_{A},0'_{A};g_{B},0_{B}\rangle\nonumber \\
 & \quad+|\mu_{M}^{B}|^{2}\langle g_{A},0_{A};e_{B},0'_{B}|\,\hat{G}(\omega)\,|g_{A},0_{A};e_{B},0'_{B}\rangle.\label{eq: exact_dimer_response_expression}
\end{align}
with, 
\begin{align}
\langle e_{A},0'_{A};g_{B},0_{B}|\hat{G}(\omega)\left|e_{A},0'_{A};g_{B},0_{B}\right\rangle  & =\frac{G_{e,A}\Big[1-J^{2}G_{e,A,v,B}(G_{e,B}+G_{v,A,e,B})\Big]}{1-J^{2}(G_{e,A}+G_{e,A,v,B})(G_{e,B}+G_{v,A,e,B})},\\
\langle g_{A},0_{A};e_{B},0'_{B}|\hat{G}(\omega)\left|g_{A},0_{A};e_{B},0'_{B}\right\rangle  & =\frac{G_{e,B}\Big[1-J^{2}G_{v,A,e,B}(G_{e,A}+G_{e,A,v,B})\Big]}{1-J^{2}(G_{e,B}+G_{v,A,e,B})(G_{e,A}+G_{e,A,v,B})},\\
\langle e_{A},0'_{A};g_{B},0_{B}|\hat{G}(\omega)\left|g_{A},0_{A};e_{B},0'_{B}\right\rangle =\langle g_{A},0{}_{A};e_{B},0'_{B}|\hat{G}(\omega)\left|e_{A},0'_{A};g_{B},0_{B}\right\rangle  & =\frac{JG_{e,A}G_{e,B}}{1-J^{2}(G_{e,A}+G_{e,A,v,B})(G_{e,B}+G_{v,A,e,B})}.
\end{align}

\noindent where, 
\begin{align}
G_{e,A} & =\frac{1}{\omega-(\omega_{e,0,A}+\omega_{g,0,B})+i\frac{\gamma_{e,A}}{2}},\\
G_{e,B} & =\frac{1}{\omega-(\omega_{e,0,B}+\omega_{g,0,A})+i\frac{\gamma_{e,B}}{2}},\\
G_{v,A,e,B} & =\frac{1}{\omega-(\omega_{g,1,A}+\omega_{e,0,B})+i\frac{\gamma_{v,A}+\gamma_{e,B}}{2}},\\
G_{e,A,v,B} & =\frac{1}{\omega-(\omega_{e,0,A}+\omega_{g,1,B})+i\frac{\gamma_{e,A}+\gamma_{v,B}}{2}}.
\end{align}

\section{Heterodimer simulation}

We have derived exact expressions for the linear absorption spectra
of a heterodimer in Section~\ref{sec:SI}. Importantly, this framework
can be straightforwardly extended to an arbitrary number of vibronic
states on both the ground- and excited-state potential energy surfaces
(PES). Consequently, one can compute the linear absorption spectrum
of any heterodimer using this protocol, without any constraint on
the value of $J$, i.e., the coupling need not be treated perturbatively.
The only requirement is to parameterize the molecular dimer such that
all parameters entering the Hamiltonian in Eq.~2 of the main text
are determined: specifically, the energies and wavefunctions of the
vibronic states, and the dipole--dipole coupling strength $J$ appearing
in Eq.~9 of the main text. Although the linear response expressions
in Eq.~\ref{eq: exact_dimer_response_expression}\textcolor{blue}{{}
}are exact, it is illuminating to explore certain perturbative regimes
of this system, where $J$ serves as the perturbative parameter, as
we do below.
\begin{table}[h]
\centering %
\begin{tabular}{|c|c|c|}
\hline 
parameters & values in cm$^{-1}$ & values in eV\tabularnewline
\hline 
\hline 
$\omega_{e,0,A}$ & $14809$ & $1.836$\tabularnewline
\hline 
$\omega_{e,0,B}$ & $15037$ & $1.864$\tabularnewline
\hline 
$\omega_{g,1,A}-\omega_{g,0,A}$ & $350$ & $4.3\times10^{-2}$\tabularnewline
\hline 
$\omega_{g,1,B}-\omega_{g,0,B}$ & $350$ & $4.3\times10^{-2}$\tabularnewline
\hline 
$J$ & $52$ & $6.5\times10^{-3}$\tabularnewline
\hline 
$\gamma_{e,A},\gamma_{e,B}$ & $60$ & $7.5\times10^{-3}$\tabularnewline
\hline 
$\gamma_{v,A},\gamma_{v,B}$ & $2.4$ & $3\times10^{-4}$\tabularnewline
\hline 
\end{tabular}\caption{Parameters of the Chlorophyll hetero dimer Ch522-Ch520 according to
Ref.~\cite{reppert2010lowest}}
\label{tab:parameters}
\end{table}

\begin{figure}
\centering{}\includegraphics[scale=0.7]{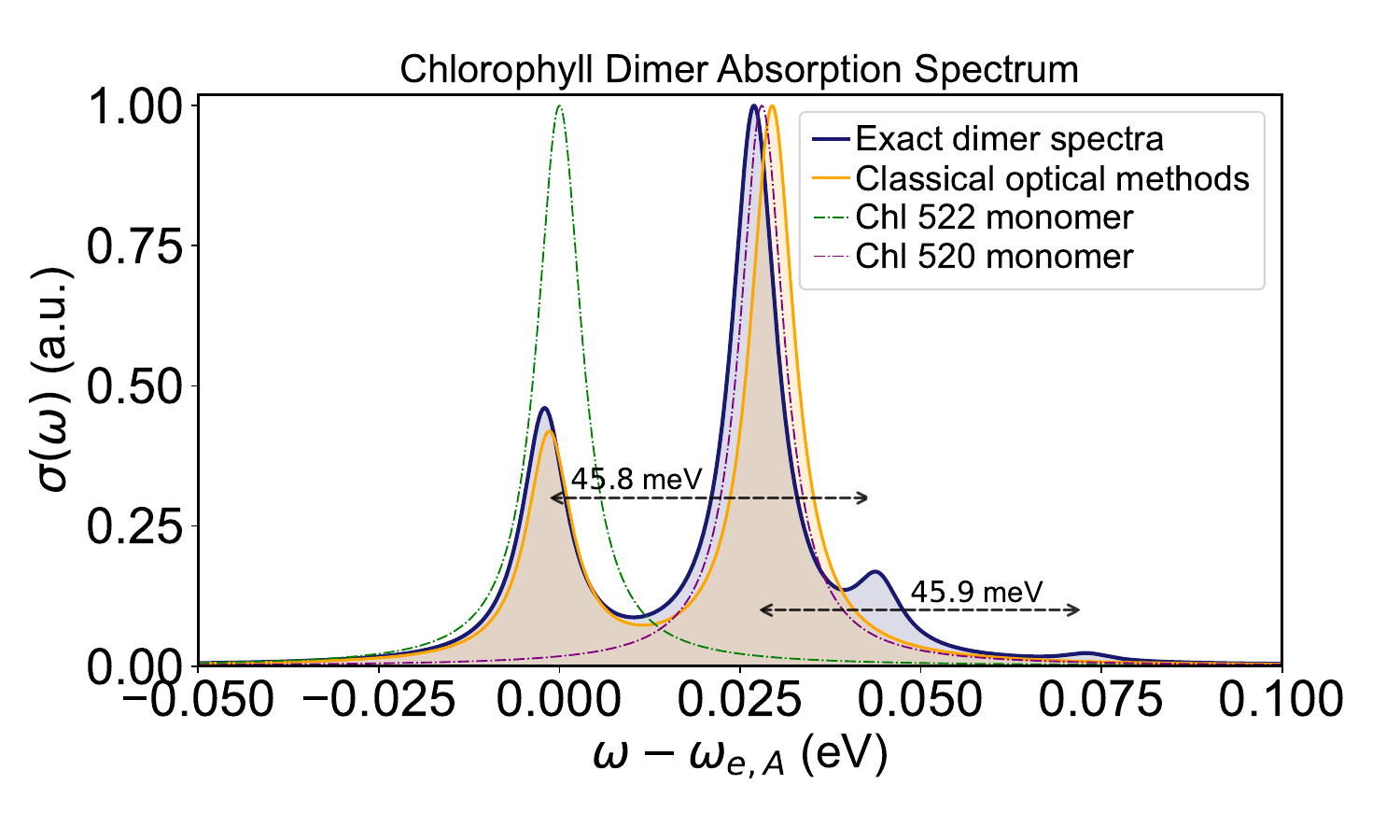}\caption{The absorption spectrum of the chlorophyll dimer (Ch522--Ch520),
as shown in Fig. 1 of the main text, is reproduced here for reference.
\label{fig: hetero_dimer_absorption_spectra}}
\end{figure}

We consider a simulation with the parameters given in Table~\ref{tab:parameters}.
These parameters correspond to the heterodimer studied in Ref.~\cite{reppert2010lowest},
where monomer A corresponds to Chl522 and monomer B corresponds to
Chl520. Specifically, the site energies are taken from Table~3, the
coupling $J$ from Table~1, and the vibrational frequencies from
Table~2 of Ref.~\cite{reppert2010lowest}. Since Table~2 contains
a large number of vibrational levels, we retain only a single representative
mode in order to remain within the three-level system framework of
the derived expressions in Eq.~\ref{eq: exact_dimer_response_expression}.
The selected vibrational mode is among the strongest ones coupled
to the electronic transition, as reflected by its Huang--Rhys factor
in Table~2 of Ref.~\cite{reppert2010lowest}. The electronic linewidth,
modeled as a Lorentzian, is taken to be $60~\text{cm}^{-1}$, as stated
in the second paragraph below Eq.~(7) of Ref.~\cite{reppert2010lowest}.
The vibrational linewidth is taken to be $3\times10^{-4}$~eV (\cite{raghavan2025high}).
Further, for simplicity, we have taken $|\mu_{M}^{A}|=|\mu_{M}^{B}|=1$.
The corresponding absorption spectrum is given in Fig. \ref{fig: hetero_dimer_absorption_spectra}.

Here, the absorption peak of monomer A has an associated Raman (of
monomer B) sideband that is shifted by approximately the vibrational
gap $\omega_{\nu,B}=\omega_{g,1,B}-\omega_{g,0,B}=0.043\text{ eV}$,
and similarly, the absorption peak of monomer B has an associated
Raman sideband shifted by approximately $\omega_{\nu,A}=\omega_{g,1,A}-\omega_{g,0,A}=0.043\text{ eV}$.
Thus, in this parameter regime, Raman-type information is directly
encoded in the linear absorption spectrum of the heterodimer. This
is consistent with the intuition developed from the ladder diagrams
presented in Section~\ref{sec:SI}: nonlinear information, specifically,
the Raman contribution (which is part of the third-order susceptibility,
$\chi^{(3)}(\omega)$), is embedded in the linear absorption spectrum
of the molecular dimer.

Here, $\left|\frac{J}{\omega_{e,0,A}-\omega_{e,0,B}}\right|\approx0.2$,
placing the heterodimer in a regime where intermolecular coupling
can be treated perturbatively. Thus, the spectra in Fig.~1 of the
main text can be rationalized by expanding Eq.~\ref{eq: exact_dimer_response_expression}
in powers of $J$. Here, the Raman-type information appears in the
diagonal components (see Section~\ref{subsec:diagonal_terms}) at
order $J^{2}$ (see Eq.~\ref{eq:second_order-1}), and in the off-diagonal
components (see Section~\ref{subsec:Off-diagonal-terms}) at order
$J^{3}$ (see Eq.~\ref{eq: third_order_off_diagonal_term}). This
is explicitly shown below: 
\begin{align}
\langle e_{A},0'_{A};g_{B},0_{B}|\hat{G}(\omega)|e_{A},0'_{A};g_{B},0_{B}\rangle & =G_{e,A}+J^{2}G_{e,A}^{2}(G_{e,B}+G_{v,A,e,B})+\mathcal{O}(J^{4})\label{eq:diagonal_A_J^2_term}\\
\langle g_{A},0_{A};e_{B},0'_{B}|\hat{G}(\omega)|g_{A},0_{A};e_{B},0'_{B}\rangle & =G_{e,B}+J^{2}G_{e,B}^{2}(G_{e,A}+G_{e,A,v,B})+\mathcal{O}(J^{4})\label{eq:diagonal_B_J^2_term}\\
\langle e_{A},0'_{A};g_{B},0_{B}|\hat{G}(\omega)|g_{A},0_{A};e_{B},0'_{B}\rangle & =\langle g_{A},0_{A};e_{B},0'_{B}|\hat{G}(\omega)|e_{A},0'_{A};g_{B},0_{B}\rangle=JG_{e,A}G_{e,B}\nonumber \\
 & \quad+J^{3}G_{e,A}G_{e,B}(G_{e,A}+G_{e,A,v,B})(G_{e,B}+G_{v,A,e,B})\nonumber \\
 & \quad+\mathcal{O}(J^{5})\label{eq:offdiag_J^3_term}
\end{align}

Thus, truncating the expansion consistently at $\mathcal{O}(J^{2})$,
the spectrum follows from Eq.~\ref{eq: exact_dimer_response_expression}.
The diagonal Green\textquoteright s function elements are given by
Eqs.~\ref{eq:diagonal_A_J^2_term} and \ref{eq:diagonal_B_J^2_term},
while the off-diagonal contribution (Eq.~\ref{eq:offdiag_J^3_term})
enters only through its leading $\mathcal{O}(J)$ term, 
\begin{align}
\langle e_{A},0'_{A};g_{B},0_{B}|\hat{G}(\omega)|g_{A},0_{A};e_{B},0'_{B}\rangle & =\langle g_{A},0_{A};e_{B},0'_{B}|\hat{G}(\omega)|e_{A},0'_{A};g_{B},0_{B}\rangle\\
 & =JG_{e,A}G_{e,B}.
\end{align}

The diagonal components contain zeroth-order terms $G_{e,A}$ and
$G_{e,B}$ (see Eq. \ref{eq:diagonal_A_J^2_term} and \ref{eq:diagonal_B_J^2_term}),
which reproduce the absorption features of the isolated monomers A
and B, respectively. The off-diagonal terms, which are first order
in $J$, contribute additional oscillator strengths at the electronic
transition frequencies of both monomers, as can be seen from the poles
of $JG_{e,A}G_{e,B}$.

\begin{figure}
\centering{}\includegraphics[scale=0.4]{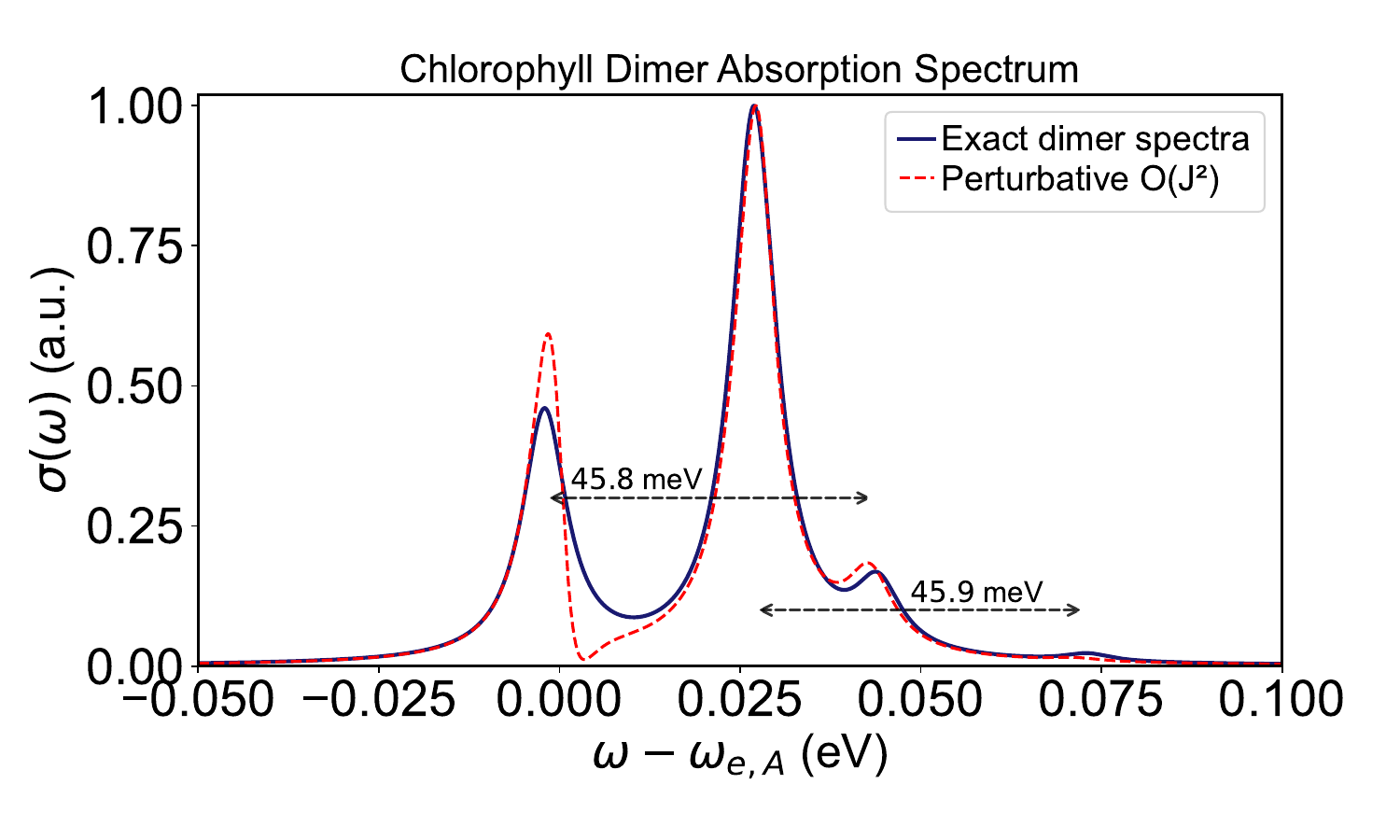}\includegraphics[scale=0.4]{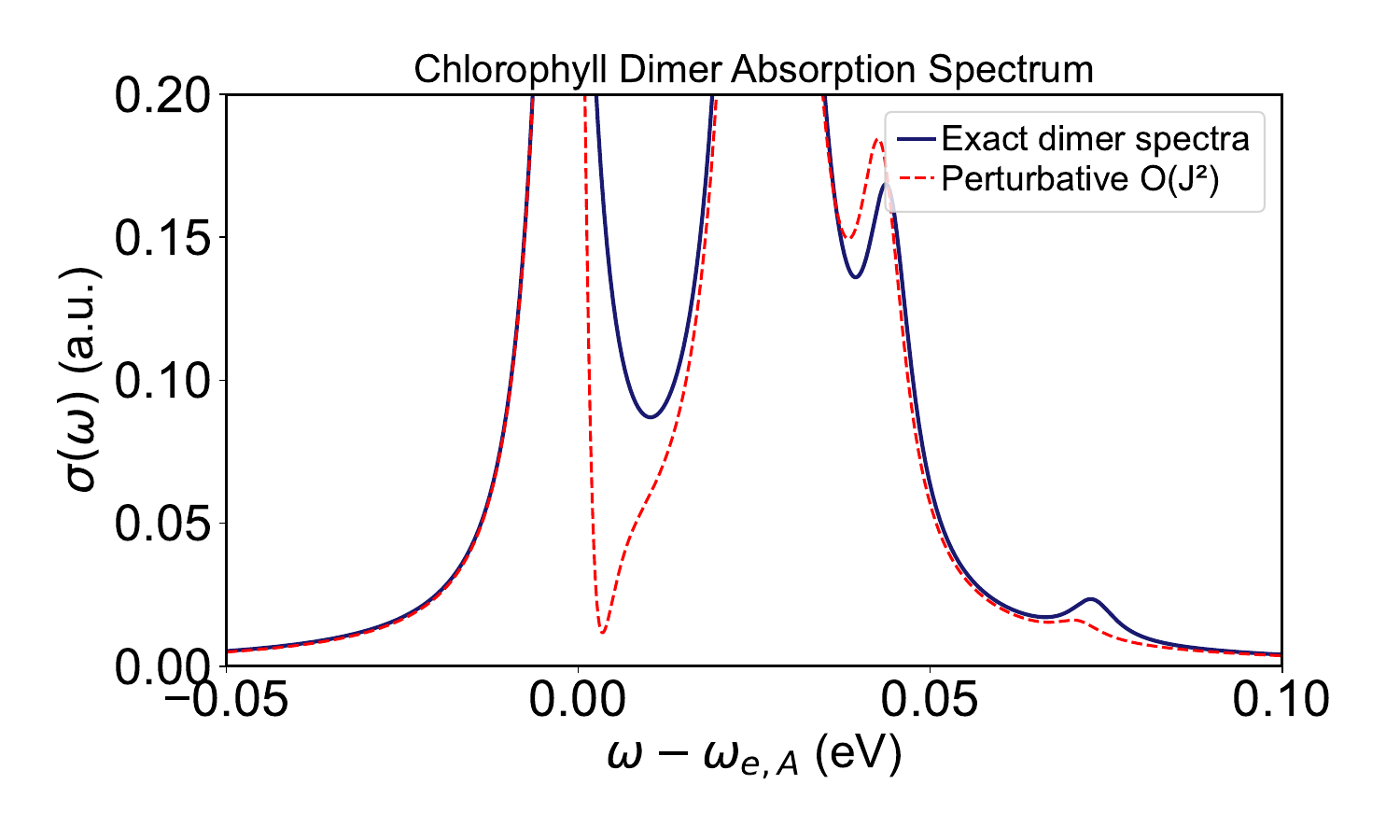}\caption{Perturbative analysis of the heterodimer spectra. The blue curve denotes
the exact dimer spectrum, corresponding to Fig.~1 of the main text,
while the red dashed curve is obtained by retaining terms only through
$\mathcal{O}(J^{2})$ in Eq.~(\ref{eq: exact_dimer_response_expression}).
As shown in the left panel, this perturbative correction is sufficient
to reproduce the Raman feature near $\omega-\omega_{e,A}\approx0.05$.
The right panel, which provides a zoomed-in view of the left panel,
further shows that the same $\mathcal{O}(J^{2})$ approximation captures
the Raman feature near $\omega-\omega_{e,A}\approx0.75$.\label{fig: perturbative_hetero_dimer_analysis}}
\end{figure}

Turning now to the central result, we examine the $\mathcal{O}(J^{2})$
terms: $J^{2}G_{e,A}^{2}(G_{e,B}+G_{v,A,e,B})$ and $J^{2}G_{e,B}^{2}(G_{e,A}+G_{e,A,v,B})$.
The first parts of these expressions, $J^{2}G_{e,A}^{2}G_{e,B}$ and
$J^{2}G_{e,A}G_{e,B}^{2}$, contribute poles at the bare electronic
transition frequencies, thereby adding oscillator strength around
the monomer peaks (also introducing some shifts) in this perturbative
regime. The second parts, $J^{2}G_{e,A}^{2}G_{v,A,e,B}$ and $J^{2}G_{e,B}^{2}G_{e,A,v,B}$,
are more significant: they generate poles shifted by the vibrational
energy gaps of the coupled (other) monomer. Specifically, $J^{2}G_{e,A}^{2}G_{v,A,e,B}$
creates a pole at $\omega=\omega_{e,0,B}+(\omega_{g,1,A}-\omega_{g,0,A})$,
as seen in Fig.~\ref{fig: hetero_dimer_absorption_spectra}, and
similarly, $J^{2}G_{e,B}^{2}G_{e,A,v,B}$ creates a pole at $\omega=\omega_{e,0,A}+(\omega_{g,1,B}-\omega_{g,0,B})$,
also seen in Fig.~\ref{fig: hetero_dimer_absorption_spectra}. These
perturbative analyses are shown in Fig. \ref{fig: perturbative_hetero_dimer_analysis}
for reference. Extending this analysis, the entire Raman spectrum
of the embedded monomers can in principle be read off from the linear
absorption spectrum of the heterodimer, in the spirit of Ref.~\cite{raghavan2025high}.

Finally, we note that in Fig.~\ref{fig: hetero_dimer_absorption_spectra},
the Raman sideband features are not shifted by exactly the vibrational
frequency; there are additional small shifts present. We attribute
these to higher-order contributions of $\mathcal{O}(J^{4})$ not retained
in our perturbative analysis, which can be systematically computed
for any system of interest.

Further, the asymmetry in the absorption peak heights (compared to
the uncoupled monomer peak heights in Fig. \ref{fig: hetero_dimer_absorption_spectra})
is a direct fingerprint of the \textit{sign} of the coupling strength
$J$. Since $J$ is positive, even its perturbative influence is sufficient
to imprint a bright--dark character onto the two eigenstates of the
dimer, a phenomenon similar to what is observed in molecular H- and
J-aggregates. To see why, recall that the dipole--field interaction
(the $\hat{\boldsymbol{\mu}}\cdot\mathbf{E}$ term embedded in Eq.
8 of the main text) selectively drives transitions to states whose
transition dipole moments add constructively. When $J>0$, it is the
higher-energy eigenstate that acquires the dominant bright character.
This superposition, in which the individual monomer transition dipoles
combine \textit{in-phase}, leads to increased oscillator strength
near the electronic transition frequency of monomer~$B$. The lower-energy
eigenstate, by contrast, receives a comparatively smaller component
of the total oscillator strength, which is why the absorption feature
near the transition frequency of monomer~$A$ appears diminished.
This logic inverts cleanly when $J<0$: the bright and dark characters
exchange roles, the oscillator strength migrates to the lower-energy
eigenstate, and the absorption peak near the transition frequency
of monomer~$A$ would instead dominate.

In summary, we have established a protocol by which the linear absorption
spectrum of a heterodimer directly encodes the Raman features of its
constituent monomers.

\section{Linear response of linear molecular aggregates}

\noindent In this section, we discuss the calculations performed on
the linear response of linear molecular aggregates that is presented
in the manuscript in Fig. (2). In order to obtain the difference between
the CPA features and the Raman features, we have compared the CPA
spectra for the linearly coupled molecular aggregate system with the
parameters shown in Table (\ref{Tab: parameters_manuscript_fig})
with the exact spectra computed for aggregates made up of $N=10$
monomers. Further, the vibrational degrees of freedom of the molecules
are modeled through shifted harmonic oscillator model~\cite{hestand2018expanded},
leading to the following Hamiltonian, 
\[
H_{e}=\omega_{v}\sum_{n}b_{n}^{\dagger}b_{n}+\omega_{v}\lambda\sum_{n}(b_{n}^{\dagger}+b_{n})\left|n\right\rangle \left\langle n\right|+\sum_{n}J\Big(\left|n+1\right\rangle \left\langle n\right|+\left|n\right\rangle \left\langle n+1\right|\Big)+\omega_{e}+\lambda^{2}\omega_{v},
\]
where, we have considered the monomers to be interacting only with
its nearest neighbors. Here, $\omega_{e}$ is the monomer electronic
excitation energy with $\lambda^{2}$ being the Huang-Rhy's factor
modeling the vibronic coupling with the harmonic vibration characterized
by $\omega_{v}$. Further, $b_{n}$ is the annihilation operator of
the vibrational mode and $\left|n\right\rangle $ is the state corresponding
to the electronic excitation only in the $n^{\text{th}}$ monomer.
Finally, the linear absorption spectra of this system is computed
using 
\[
\sigma(\omega)\propto-\Im\left\langle G\right|\mu G(\omega)\mu\left|G\right\rangle ,
\]
where, $G(\omega)$ is the Green's function of the whole system, $\mu$
is the total dipole moment operator and $\left|G\right\rangle $ represents
the state of the system where all the monomers are in the global ground
state.

\begin{table}
\centering{}%
\begin{tabular}{|c|c|}
\hline 
Parameters & value\tabularnewline
\hline 
\hline 
$\lambda^{2}$ & $1$ eV\tabularnewline
\hline 
$\omega_{v}$ & $1$ eV\tabularnewline
\hline 
$J$ & $-3$ eV\tabularnewline
\hline 
$\gamma$ & $0.18$ eV\tabularnewline
\hline 
$N$ & $10$\tabularnewline
\hline 
$M$ & $2$\tabularnewline
\hline 
\end{tabular}\caption{Parameters corresponding to the linear absorption spectra of linear
aggregates shown in Fig. (5) of the manuscript \label{Tab: parameters_manuscript_fig}}
\end{table}

\section{Many particle approximations}

\begin{figure}

\begin{centering}
\begin{tabular}{|c|c|}
\hline 
\textbf{Parameters}  & \textbf{Value} \tabularnewline
\hline 
\hline 
$N$  & $6$ \tabularnewline
\hline 
$\omega_{0}$  & $1\text{eV}$ \tabularnewline
\hline 
$\lambda^{2}$  & $1$ \tabularnewline
\hline 
$J_{n,n+1}$  & $-1\text{eV}$ \tabularnewline
\hline 
$M$  & $2$ \tabularnewline
\hline 
$\gamma$  & $0.18\text{eV}$ \tabularnewline
\hline 
\end{tabular}
\par\end{centering}
\centering{}\includegraphics[width=0.5\linewidth]{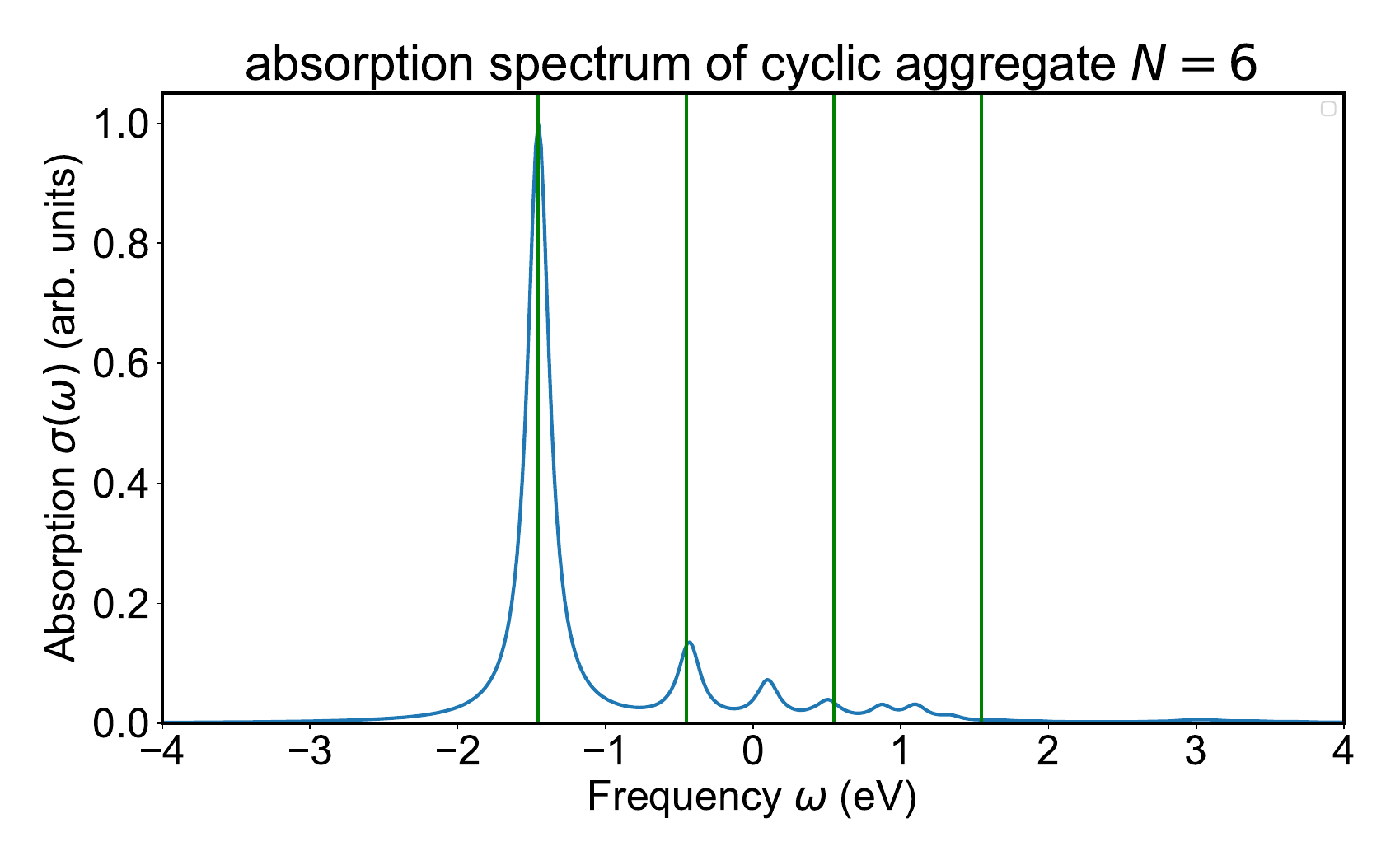}
\caption{Parameters and the corresponding exact absorption spectra for an idealized
circular aggregates, where, all of the dipole moments are aligned.
The parameters corespond to the Fig. 17 (i) in Hestand et al.~\cite{hestand2018expanded}.
We have used eV for the purpose of the simulation, however, the unit
doesn't affect the spectra.}
\label{Fig: TPA_fig_exact} 
\end{figure}

\noindent In addition to Fig.~(2) shown in the manuscript, where
we interpret the corrections to the CPA in the molecular aggregate
spectra as Raman signatures, we also explore in this section the features
that are captured by the Two-Particle Approximation (TPA). In Fig.~(\ref{Fig: TPA_fig_exact}),
we show the exact simulation of the nearest-neighbor--coupled molecular
aggregate system explored in the literature (Fig.~17(i) in Hestand
et al.~\cite{hestand2018expanded}). We observe Raman-type signatures
appearing in the strongly coupled regime, similar to Fig.~(2) in
the manuscript, which are captured by the TPA. Although the exact
spectra capture finer features than the TPA, the Raman-type features
already appear within the TPA. This is not entirely surprising, as
the two-particle states in the TPA include configurations where monomers
host vibrational excitations in the ground electronic state. 

\end{document}